\documentclass[preprint]{aastex61}

\newcommand {\lir} {L_{\rm IR}}

\shorttitle{STUDIES: Faint  450\,$\mu$m Counts}
\shortauthors{Wang et al.}

\begin{document}

\title{SCUBA-2 Ultra Deep Imaging EAO Survey (STUDIES): Faint-End Counts at 450 \MakeLowercase{$\mu$m}}%---a Single Power Law between 1 and 25 \MakeLowercase{m}J\MakeLowercase{y}}

\author{Wei-Hao Wang} 
\affiliation{Academia Sinica Institute of Astronomy and Astrophysics (ASIAA), No.\ 1, Sec.\ 4, Roosevelt Rd., Taipei 10617, Taiwan}
\email{whwang@asiaa.sinica.edu.tw}

\author{Wei-Ching Lin}
\affiliation{Institute of Physics, National Taiwan University, Taipei 10617, Taiwan}
\affiliation{Academia Sinica Institute of Astronomy and Astrophysics (ASIAA), No.\ 1, Sec.\ 4, Roosevelt Rd., Taipei 10617, Taiwan}

\author{Chen-Fatt Lim}
\affiliation{Institute of Astrophysics, National Taiwan University, Taipei 10617, Taiwan}
\affiliation{Academia Sinica Institute of Astronomy and Astrophysics (ASIAA), No.\ 1, Sec.\ 4, Roosevelt Rd., Taipei 10617, Taiwan}

\author{Ian Smail}
\affiliation{Centre for Extragalactic Astronomy, Department of Physics, Durham University, Durham DH1 3LE, UK}

\author{Scott C.\ Chapman}
\affiliation{Department of Physics and Atmospheric Science, Dalhousie University, Halifax, NS B3H 3J5 Canada}

\author{Xian Zhong Zheng}
\affiliation{Purple Mountain Observatory, Chinese Academy of Sciences, Nanjing 210008, China}

\author{Hyunjin Shim}
\affiliation{Department of Science Education, Kyungpook National University, Bukgu, Daegu 41566, Korea}

\author{Tadayuki Kodama}
\affiliation{National Astronomical Observatory of Japan (NAOJ), Mitaka, Tokyo 181-8588, Japan}

\author{Omar Almaini}
\affiliation{School of Physics and Astronomy, University of Nottingham, Nottingham NG7 2RD, UK}

\author{Yiping Ao}
\affiliation{National Astronomical Observatory of Japan (NAOJ), Mitaka, Tokyo 181-8588, Japan}
\affiliation{Purple Mountain Observatory, Chinese Academy of Sciences, Nanjing 210008, China}

\author{Andrew W.\ Blain}
\affiliation{Department of Physics and Astronomy, University of Leicester, Leicester LE1 7RH, UK}

\author{Nathan Bourne}
\affiliation{Institute for Astronomy, University of Edinburgh, Royal Observatory, Edinburgh EH9 3HJ, UK}

\author{Andrew J.\ Bunker}
\affiliation{Department of Physics, University of Oxford, Oxford OX13RH, UK}
\affiliation{Kavli Institute for the Physics and Mathematics of the Universe, Kashiwa, 277-8583, Japan}

\author{Yu-Yen Chang}
\affiliation{Academia Sinica Institute of Astronomy and Astrophysics (ASIAA), No.\ 1, Sec.\ 4, Roosevelt Rd., Taipei 10617, Taiwan}

\author{Dani C.-Y.\ Chao}
\affiliation{Institute of Astronomy, National Tsing Hua University, Hsinchu, Taiwan. 30013}
\affiliation{Current address: Max Planck Institute for Astrophysics, D-85740 Garching, Germany}

\author{Chian-Chou Chen}
\altaffiliation{ESO Fellow}
\affiliation{European Southern Observatory, Garching, Germany}

\author{David L.\ Clements}
\affiliation{Astrophysics Group, Imperial College London, Blackett Laboratory, London SW7 2AZ, UK}

\author{Christopher J.\ Conselice}
\affiliation{School of Physics and Astronomy, University of Nottingham, Nottingham NG7 2RD, UK}

\author{William I.\ Cowley}
\affiliation{Kapteyn Astronomical Institute, University of Groningen, Groningen, the Netherlands}

\author{Helmut Dannerbauer}
\affiliation{Instituto de Astrof\'{i}sica de Canarias (IAC), Tenerife, Spain}
\affiliation{Universidad de La Laguna, Dpto. Astrofisica, Tenerife, Spain}

\author{James S.\ Dunlop}
\affiliation{Institute for Astronomy, University of Edinburgh, Royal Observatory, Edinburgh EH9 3HJ, UK}

\author{James E.\ Geach}
\affiliation{Centre for Astrophysics Research, School of Physics, Astronomy and Mathematics, University of Hertfordshire, Hatfield AL10 9AB, UK}

\author{Tomotsugu Goto}
\affiliation{Institute of Astronomy, National Tsing Hua University, Hsinchu, Taiwan. 30013}

\author{Linhua Jiang}
\affiliation{Kavli Institute for Astronomy and Astrophysics, Peking University, Beijing 100871, China}

\author{Rob J.\ Ivison}
\affiliation{European Southern Observatory, Garching, Germany}
\affiliation{Institute for Astronomy, University of Edinburgh, Royal Observatory, Edinburgh EH9 3HJ, UK}

\author{Woong-Seob Jeong}
\affiliation{Korea Astronomy and Space Science Institute, Yuseong-gu, Daejeon 34055, Korea}
\affiliation{University of Science and Technology, Yuseong-gu, Daejeon 34113, Korea}

\author{Kotaro Kohno}
\affiliation{Institute of Astronomy, University of Tokyo, Mitaka, Tokyo 181-0015, Japan}
\affiliation{Research Center for the Early Universe, University of Tokyo, Bunkyo, Tokyo 113-0033, Japan}

\author{Xu Kong}
\affiliation{Department of Astronomy, University of Science and Technology of China, Hefei, Anhui 230026, China}

\author{Chien-Hsu Lee}
\affiliation{Subaru Telescope, NAOJ, Hilo, HI 96720, USA}

\author{Hyung Mok Lee}
\affiliation{Department of Physics and Astronomy, Seoul National University, Seoul 08826, Korea}

\author{Minju Lee}
\affiliation{National Astronomical Observatory of Japan (NAOJ), Mitaka, Tokyo 181-8588, Japan}
\affiliation{Department of Astronomy, Graduate School of Science, The University of Tokyo, Bunkyo, Tokyo 113-0033, Japan}

\author{Micha{\l} J.\ Micha{\l}owski}
\affiliation{Astronomical Observatory Institute, Faculty of Physics, Adam Mickiewicz University, 60-286 Pozna{\'n}, Poland}

\author{Iv\'{a}n Oteo}
\affiliation{Institute for Astronomy, University of Edinburgh, Royal Observatory, Edinburgh EH9 3HJ, UK}
\affiliation{European Southern Observatory, Garching, Germany}

\author{Marcin Sawicki}
\affiliation{Saint Mary's University, Department of Astronomy \& Physics, Halifax, B3J 3Z4, Canada}

\author{Douglas Scott}
\affiliation{Department of Physics and Astronomy, University of British Columbia, Vancouver, BC V6T 1Z1, Canada}

\author{Xin Wen Shu}
\affiliation{Department of Physics, Anhui Normal University, Wuhu, Anhui, 241000, China}

\author{James M.\ Simpson}
\altaffiliation{EACOA Fellow}
\affiliation{Academia Sinica Institute of Astronomy and Astrophysics (ASIAA), No.\ 1, Sec.\ 4, Roosevelt Rd., Taipei 10617, Taiwan}

\author{Wei-Leong Tee}
\affiliation{Institute of Physics, National Taiwan University, Taipei 10617, Taiwan}
\affiliation{Academia Sinica Institute of Astronomy and Astrophysics (ASIAA), No.\ 1, Sec.\ 4, Roosevelt Rd., Taipei 10617, Taiwan}

\author{Yoshiki Toba}
\affiliation{Academia Sinica Institute of Astronomy and Astrophysics (ASIAA), No.\ 1, Sec.\ 4, Roosevelt Rd., Taipei 10617, Taiwan}

\author{Elisabetta Valiante}
\affiliation{School of Physics and Astronomy, Cardiff University, Cardiff CF24 3AA, UK}

\author{Jun-Xian Wang}
\affiliation{Department of Astronomy, University of Science and Technology of China, Hefei, Anhui 230026, China}

\author{Ran Wang}
\affiliation{Kavli Institute for Astronomy and Astrophysics, Peking University, Beijing 100871, China}

\author{Julie L.\ Wardlow}
\affiliation{Centre for Extragalactic Astronomy, Department of Physics, Durham University, Durham DH1 3LE, UK}

\begin{abstract}
The SCUBA-2 Ultra Deep Imaging EAO Survey (STUDIES) is a three-year JCMT Large Program aiming at reaching the 
 450\,$\mu$m confusion limit in the COSMOS-CANDELS region, to study a 
representative sample of the high-redshift far-infrared galaxy population that gives rise to the bulk of the far-infrared background.  
We present the first-year data from STUDIES.  
We have reached a 450\,$\mu$m noise level of 0.91~mJy for point sources at the map center, covered an area of 151~arcmin$^2$, 
and detected 98 and 141 sources at 4.0 and 3.5 $\sigma$, respectively.  Our derived counts are best constrained in the
3.5--25~mJy regime using directly detected sources. 
Below the detection limits, our fluctuation analysis further constrains the slope of the counts down to 1 mJy.
The resulting counts at 1--25 mJy are consistent with a power law
having a slope of $-2.59$ ($\pm0.10$ for 3.5--25~mJy, and $^{+0.4}_{-0.7}$ for 1--3.5 mJy). 
There is no evidence of a faint-end termination or turn-over of the counts in this flux density range.  
Our counts are also consistent with previous SCUBA-2 blank-field and lensing cluster surveys.
The integrated surface brightness from our counts down to 1 mJy is $90.0\pm17.2$ 
Jy deg$^{-2}$, which can account for up to $83^{+15}_{-16}\%$ of the \emph{COBE} 450\,$\mu$m background.
We show that \emph{Herschel} counts at 350 and 500\,$\mu$m are significantly higher than our 450\,$\mu$m counts, likely caused by its large beam and source clustering.
%Analyses of high-redshift galaxies purely based on \emph{Herschel}-SPIRE data may therefore contain biases. 
High-angular resolution instruments like SCUBA-2 at 450\,$\mu$m are therefore highly beneficial for measuring the luminosity and 
spatial density of high-redshift dusty galaxies.
\end{abstract}


\section{Introduction}

Since the advent of the Submillimeter Common User Bolometer Array \cite[SCUBA,][]{holland99}
on the James Clerk Maxwell Telescope (JCMT) 
and the discovery of the first submillimeter galaxies \citep[SMGs,][]{smail97,hughes98,barger98,eales99} two decades ago,
tremendous progress has been made in understanding this important dusty galaxy population
(see reviews by \citealt{blain02,casey14}).
Wide-field 850\,$\mu$m surveys made with SCUBA and other bolometer array cameras have provided large
samples of SMGs, while interferometric followup observations have enabled counterpart
identification and detailed studies.

We now know that 850\,$\mu$m-selected SMGs are ultraluminous ($\lir > 10^{12}$~L$_\sun$)
galaxies at $z\simeq1$--6. Their redshift distribution peaks at $z\simeq2.5$ \citep{chapman03a,chapman05,barger14,simpson14,chen16,michalowski17} with
a $z>4$ high-redshift tail \citep[e.g,][]{younger07,wang07,dannerbauer08,shu16,asboth16,ivison16}.  The exact redshift distribution of SMGs also strongly depends
on the selection waveband; longer wavelengths may pick up SMGs at higher redshifts 
\citep[e.g.,][]{chapin09,roseboom13,zavala14,strandet16}.  X-ray and infrared studies of 850\,$\mu$m SMGs show that a modest fraction about 20\% of them 
host active galactic nuclei (AGN) but most of them have total infrared luminosities dominated by star formation
\citep{almaini03,alexander05,valiante07,pope08,menendez09,laird10,wang13}.  Their star-formation rates (SFRs) are typically $10^2$ to $>10^3$~M$_\sun$~yr$^{-1}$,
and they contribute a significant fraction ($\simeq30\%$) of the total SFR density at $z=1$--5 \citep[e.g.,][]{barger00,barger12,chapman05,wang06,michalowski10,wardlow11,casey13,swinbank14,cowie17,bourne17}.  
On the other hand, it remains unclear whether the extremely large SFRs are triggered by major mergers or by disk instabilities.
Results from multi-wavelength morphological studies have not converged \citep{conselice03,chapman03a,swinbank10,swinbank11,chen15,simpson15,hodge16,dunlop17},
and models that favor different mechanisms exist \citep{narayanan10,narayanan15,hayward11,hayward12,lacey16}.

It is important to point out that attempts to understand the SMG population has been overwhelmingly focused on the
bright end at 850\,$\mu$m. The poor angular resolution of single-dish telescopes (e.g., FWHM~$=15\arcsec$ for JCMT at 850\,$\mu$m) 
produces a relatively bright limit due to source blending ($S_{850}\sim2$--3 mJy).
This effect is known as ``confusion'' \citep{condon74}; it prevents 
the detection of faint sources and the full resolution of the extragalactic background light (EBL).  
In the millimeter and submillimeter (mm/submm) bands, sources detected in single-dish confusion-limited blank-field 
surveys typically account for only 10--40\% of the EBL 
\citep{barger99,borys03,greve04,wang04,coppin06,weiss09,scott10,hatsukade11,geach17},
and the bulk of the EBL remains unresolved.
Surveys in strong lensing cluster fields can nearly fully resolve the mm/submm EBL 
\citep{cowie02,smail02,knudsen08,johansson11,chen13b,hsu16} and provide insight into the nature of the faint sources 
\citep{chen14,hsu16}.  However, the sample sizes of such strongly lensed sources remain small, and source-plane expansion 
and magnification bias may make cosmic variance a stronger effect in lensing-cluster surveys.

Recently, ALMA has realized its tremendous sensitivity. Its high resolution makes it essentially confusion-free and able to 
detect sources that comprise the bulk of the mm/submm EBL.
However, because of the small primary beam of its antennas, unbiased ALMA surveys are only 
able to image a few arcmin$^2$ to substantial depths \citep{umehata15,hatsukade16,aravena16,dunlop17}.
Larger samples of fainter sources have been serendipitously detected in ALMA archival data \citep{hatsukade13,ono14,carniani15,fujimoto16} including those
in the ALMA calibration fields \citep{oteo16}, but biases caused by clustering and cosmic variance on these small scales are a potential concern.
 
The 450\,$\mu$m window for deep submillimeter surveys was truly opened up by SCUBA-2 \citep{holland13}.
At 450\,$\mu$m, the nearly 2 times increase in angular resolution from 850\,$\mu$m makes SCUBA-2 much
less confusion limited.  This enables the direct detection of fainter SMGs.  This also makes multi-wavelength
counterpart identification less ambiguous.  Previous 450\,$\mu$m SCUBA-2 surveys reached 
sensitivities (1 $\sigma$ noise) of $\gtrsim1$ to 10 mJy  in various
blank fields and in lensing clusters.  Among these, the deepest blank-field surveys \citep{geach13,zavala17} resolved 20--30\% of the 
450\,$\mu$m EBL into point sources down to 6 mJy.  \citet{geach13} also estimated that the confusion noise of 
JCMT/SCUBA-2 at 450\,$\mu$m is approximately 1 mJy, but a detection limit of a comparable flux has not been reached by any blank-field surveys.
On the other hand, lensing-cluster surveys \citep{chen13a,chen13b,hsu16} had reached intrinsic
fluxes of $\sim1$ mJy on a smaller sample of highly magnified sources, and nearly fully resolved the EBL.
For comparison, confusion limited \emph{Herschel} SPIRE imaging (FWHM $\simeq30\arcsec$ at 500\,$\mu$m) can 
detect sources brighter than around 20 mJy at 250, 350, and 500\,$\mu$m \citep{glenn10,clements10,oliver10,valiante16}, and these sources 
account for only about 15\% of the FIR EBL at these wavelengths.  Unlike the 850\,$\mu$m and millimeter bands, the 450\,$\mu$m band has weaker negative
$K$-correction and is less sensitive to $z>4$ galaxies, because the
peak of the redshifted dust spectral energy distribution (SED) shifts out of the passband. However, for dust temperature of a few tens of Kelvin,
the 450\,$\mu$m band probes the SED peak of galaxies at $z\simeq1$--2, the peak epoch of both cosmic star formation and quasar activity. 
There is clearly great potential for SCUBA-2 to reach much deeper 
than \emph{Herschel} at 450\,$\mu$m, and to detect the faint dusty galaxies that give rise to the bulk of the far-infrared (FIR) EBL
and hence the bulk of the cosmic star formation. 

To exploit the potential of SCUBA-2, we initiated a new program, the SCUBA-2 Ultra Deep Imaging EAO (East-Asian Observatory) 
Survey (STUDIES; program ID: M16AL006).  It is one of the new JCMT Large Programs under the operation of EAO, starting from late 2015.  
Its goal is to map a $R\sim8\arcmin$ region and get close to the confusion limit at 450\,$\mu$m in three years with 330 hrs of observations.  
The survey field is at the center of the Cosmic Evolution Survey \citep[COSMOS,][]{scoville07} within the area of the
Cosmic Assembly Near-infrared Deep Extragalactic Legacy Survey \citep[CANDELS,][]{grogin11,koekemoer11}.
The STUDIES field is also within the wider and shallower SCUBA-2 survey of \citet{casey13}. The southern shallower region of STUDIES 
overlaps with the 450\,$\mu$m COSMOS pointing of the SCUBA-2 Cosmology Legacy Survey (S2CLS, \citealt{geach13}). 
STUDIES thus takes advantage of the extremely rich multi-wavelength data in the COSMOS-CANDELS region and the
heritage of previous SCUBA-2 surveys.

In its first year of observations, STUDIES had accumulated approximately 40\% of the data and reached an r.m.s.\ noise of 0.91 mJy at 
the map center at 450\,$\mu$m.  The image has a diameter of $15\arcmin$. The full 3-yr STUDIES image will be $1.6\times$ deeper 
over the same area.  The parallel 850\,$\mu$m imaging will provide an image that is confusion limited over the entire area.
An early science result from STUDIES can be found in \citet{simpson17}, where we detected a ``passive'' galaxy at 450\,$\mu$m and 
demonstrated that non-detections in \emph{Herschel} bands do not rule out an active star-forming system.  Multi-wavelength properties 
of the detected galaxies will be presented in our future papers.  Because the progress in 2017 is much slower, caused by the poor 
weather and instrument servicing work, here we present the first-year STUDIES data and our improved constraints on the 450\,$\mu$m source counts.

FIR and mm/submm number counts provide sensitive tests to galaxy evolution models 
\citep[e.g.,][]{baugh05,valiante09,bethermin12b,hayward13,cowley15,lacey16,bethermin17} and require strong evolution in
the properties of FIR luminous galaxies.  The counts can be also compared with satellite EBL measurements, to examine whether there remain significant 
galaxy populations that are unaccounted for in the imaging surveys.  Ultimately, when the imaging surveys are sufficiently
deep, the integrated surface brightness from the resolved source counts can put the various satellite EBL measurements to the test. 
We derive the 450\,$\mu$m counts with both direct detections of bright sources and a fluctuation analysis for faint sources using  
the first-year STUDIES data.   
In Section~\ref{sec_data}, we describe the SCUBA-2
450\,$\mu$m observations and data reduction.  In Section~\ref{sec_extract_counts}, we derive the source counts at $S_{450}>3.5$ mJy using direct  4-$\sigma$ 
detections and with simulations.  We refer to this flux density regime as the ``bright end.'' In Section~\ref{sec_mle}, we examine the image background fluctuation below 4~$\sigma$
and use a maximum likelihood method (often referred as ``$P(D)$'' analysis) to constrain the counts below 3.5 mJy.  We refer to this as the 
``faint end.''  In Section~\ref{sec_discuss}, we compare
our counts with other counts measurements and models, and estimate the contributions to the 450\,$\mu$m EBL.  We summarize in Section~\ref{sec_summary}.
We analyze the various bias effects and source blending in our observations and source extraction in Section~\ref{sec_a1}, and verify that
our results are not biased by source clustering at the scale of our beam size in Section~\ref{sec_a2}.

\section{STUDIES 450\,$\mu$\MakeLowercase{m} Data}\label{sec_data}
\subsection{Observations}

We carried out the 450\,$\mu$m SCUBA-2 observations between December 30, 2015 and May 11, 2016, under the
best submillimeter weather conditions on Maunakea (Band 1, $\tau _{\rm 225\,GHz} < 0.05$, or precipitable water vapor, PWV $< 0.83$ mm).  
SCUBA-2 also takes 850\,$\mu$m data simultaneously, and the data will be presented elsewhere.
The opacity was actively monitored with a water vapor radiometer using the 183 GHz water line
at the telescope pointing direction \citep{dempsey12}, to ensure that all observations were carried out under Band-1 conditions.  
The median PWV is 0.615~mm and the 90th-percentile range is 0.440--0.803 mm,
corresponding to 450\,$\mu$m opacities of $\tau_{\rm 450 \mu m}=0.768$, and 0.590--0.958, respectively.
Hourly telescope pointing checks and less frequent focusing were conducted on the nearby infrared source IRC+10216.
Typical pointing offsets were within $1\arcsec$, much smaller than the 450\,$\mu$m beam.
Nightly flux calibration was obtained by imaging Uranus, Mars, and the infrared sources CRL618, CRL2688, and Arp 220.
We conducted the imaging using the ``CV Daisy'' scan 
pattern,\footnote{\url http://www.eaobservatory.org/jcmt/instrumentation/continuum/scuba-2/observing-modes/} 
which creates a circular map of $R\gtrsim6\arcmin$ with a $R\simeq1\farcm5$ deep core at the center and rapidly
increasing noise at $R>5\arcmin$.  Each scan spanned 30 minutes in time.
The scans were slightly offset from each other to even out the effects of noisy bolometers.  This also slightly expands the map to $R\simeq7\farcm5$.
The total on-source time was approximately 120 hrs, accounting for 40\% of the total allocated integration of STUDIES.

\subsection{Data Reduction}
We performed the data reduction using the Sub-Millimeter Common User Reduction Facility \citep[SMURF,][]{chapin13}
and the PIpeline for Combining and Analyzing Reduced Data \citep[PICARD,][]{jenness08}.  The time-stream data from
the SCUBA-2 bolometers contain noise and signal from the background (atmosphere and ambient thermal emission), 
as well as astronomical objects.   To extract the astronomical signal from the time streams and to map the results onto a 
celestial projection, we adopted the \texttt{Dynamic Iterative Map-Maker} (DIMM, \citealp{jenness11,chapin13}) routine of 
SMURF.   We used the standard ``blank field'' configuration file, which aims to detect extremely low signal-to-noise point 
sources from deep observations. First, flat-fields were applied to the time streams by using the flat scans that 
bracket each science observation. The flat-field procedure subtracts a polynomial baseline fit from each 
time stream and scales the data to units of pW. Next, DIMM enters an iterative stage that attempts to fit the 
data with a model comprising a common-mode background signal, astronomical signal, and noise. The iterations 
were repeated only four times, as we do not expect a single science scan to have any significant signal to well constrain 
the model.  We verified that all the final scans do not change significantly with further iterations, meaning that the residual 
between the model and data has converged.

To obtain the absolute flux scale, we measured the flux conversion factors (FCFs) from a subset of 140 standard submillimeter 
calibrators that were observed during the STUDIES campaign, after excluding extreme values (usually from early evening and morning observations).
We obtained a mean FCF of $490$ Jy beam$^{-1}$ pW$^{-1}$ and an r.m.s.\ scatter of $\pm136$ Jy beam$^{-1}$ pW$^{-1}$.
The nominal value for SCUBA-2 at 450\,$\mu$m is $491\pm67$ Jy pW$^{-1}$ \citep{dempsey13}.
Our scatter is twice larger than this and those in earlier SCUBA-2 450\,$\mu$m studies in the literature.
An about twice larger yearly scatter is also seen in the SCUBA-2 Calibration Database\footnote{\url http://www.eaobservatory.org/sc2cal} for 2016.
The large scatter in our FCF therefore indicates a real night-to-night variation, rather than problems in our data or our reduction.
However, other than the larger scatter, we do not observe a long-term trend in the mean FCF, and our mean FCF is consistent with the nominal value.
We therefore applied the value of $490$ Jy beam$^{-1}$ pW$^{-1}$ to our data.  In general, we expect a 12\% uncertainty in flux calibration
\citep{dempsey13}, which is a combination of a 5\% uncertainty in the absolute calibration of planetary models and another 10\% uncertainty
in the determination of the FCF.

%A flux conversion factor (FCF) of is applied to the 
%individual clean scans to give a map unit of flux density per beam. This FCF value was 
%derived from averaging a subset of standard submillimeter calibrators that were observed during the STUDIES campaign, after
%excluding extreme values (usually from early evening and morning observations). It is also remarkably close to the 
%canonical FCF value of 491 Jy pW$^{-1}$ \citep{dempsey13}, suggesting small measurement errors in
%the flux calibration.  

\begin{figure*}[t!]
\epsscale{1.05}
\plotone{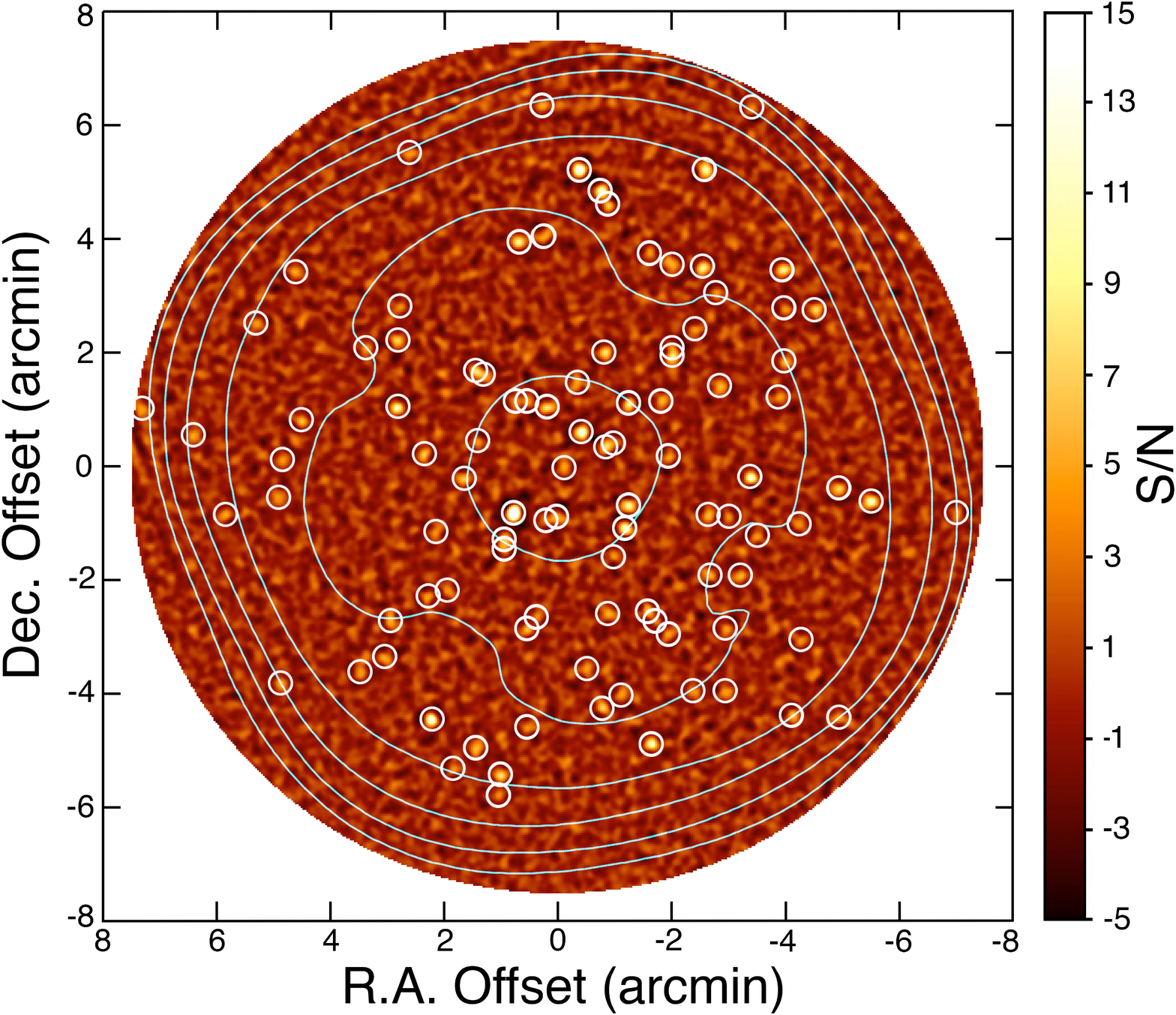}
\caption{STUDIES 450\,$\mu$m S/N map. Only the deeper $R=7\farcm5$ is shown here.  The map center is at R.A.\ = 10:00:30.2, Dec.\ = +02:26:50.  
The contours represent noise levels from 1 mJy with a multiplicative step of $\sqrt{2}$.
Sources detected at 4 $\sigma$ are marked with $R=12\arcsec$ circles; there are 98 such sources.  The easternmost source is not included in our number
counts because of the requirement  on noise of the source flux.  Several of them have nearby companions with distances of $\sim10\arcsec$,
identified by our CLEAN-like source extraction algorithm (Section~\ref{sec_extraction}).  There are an additional 43 sources detected at 3.5--4 $\sigma$. 
These fainter sources are not labeled here to avoid confusion. 
\label{fig1}} 
\end{figure*}

After each scan was reduced and flux calibrated, we adopted the \texttt{MOSAIC\_JCMT\_IMAGES} recipe from PICARD 
to combine all the products into a final map. 
To optimize the detection of point sources, we convolved the map with a Gaussian kernel that is matched to the instrumental 
point-spread function (PSF).  For this, we adopted the PICARD recipe \texttt{scuba2\_matched\_filter}, which first smooths the map 
by convolving it with a $20\arcsec$ Gaussian kernel and subtracts this smoothed map from the image to remove any large-scale structure. 
Then a normalized Gaussian kernel with a FWHM set to the diffraction limit of the telescope was used to convolve the map to obtain
the optimal point-source S/N for each pixel \citep{stetson87,serjeant03}. 
The same process was also applied to the calibrators before the FCF was measured.
Also, although the above procedure builds in a flux adjustment to compensate the flux loss caused by the subtraction of the $20\arcsec$-smoothed
image, additional adjustment may be required.  In order to see the effect on source fluxes of the blank-field configuration of DIMM and the above convolution procedures, 
we inserted artificial point sources with the instrumental FWHM into the data streams. For source flux densities from sub-Jy to a few Jy,
we found that the above reduction procedures attenuated the flux density by 6.2\%, on average. We therefore made a 6.2\% flux adjustment
to the map.  This is slightly less than the 10\% adjustment reported by \citet{geach13} and \citet{chen13b}, but the difference is 
well within the generally accepted 10\% calibration uncertainty.

\begin{figure}[t!]
%\epsscale{0.6}
\plotone{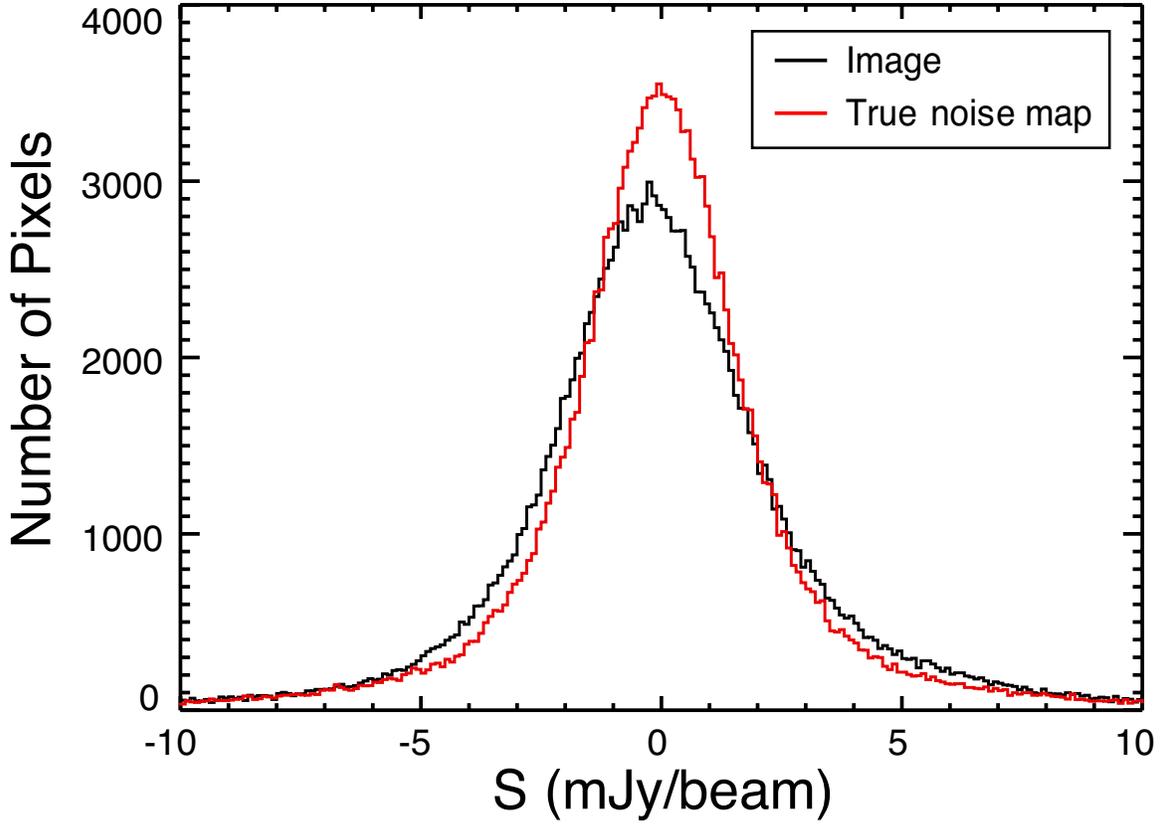}
\caption{Distribution of pixel brightness in the 450\,$\mu$m image (black) and in the ``true noise'' map (red, see Section~\ref{sec_counts}).
The excesses in the both positive and negative sides in the image are produced by celestial objects and the negative bowling in the PSF.
The noise histogram is a sum of Gaussians of various widths, because of the non-uniform sensitivity distribution.
\label{fig2}} 
\end{figure}

Our final match-filtered map has a noise at the map center of 0.91 mJy for a point source, and the noise increases to around 
10 mJy toward the map edge at a radius of $7\farcm5$.  The 0.91 mJy sensitivity is about 30\% deeper than that in the deepest
450\,$\mu$m map in the literature \citep{zavala17}, and is comparable to the deepest 450\,$\mu$m pointing in S2CLS (Geach et al., in prep.).
We present our S/N map in Fig.~\ref{fig1}, and the histograms of the pixel flux in Fig.~\ref{fig2}.
In Fig.~\ref{fig1}, we can see negative rings around the brightest sources, produced by the above match-filtering process.
The negative ring in the PSF is responsible for the negative excess in Fig.~\ref{fig2} comparing to pure noise.   
We verify the astrometry of our image by comparing the source
positions in our image and in the Very Large Array 3 GHz image \citep{smolcic17}.
Among the 98 4-$\sigma$ sources extracted from our image (Section~\ref{sec_extraction}), 59 have 3 GHz counterparts
with $4\arcsec$ search radii.  We expect 1.3 of them to be chance alignments, given the spacial density of the radio sources.
The mean positional offset between 450\,$\mu$m and 3 GHz is $0\farcs3$ along 
R.A.\ and $0\farcs2$ along Dec., and the dispersions are $\sim1\farcs5$. These are all much
smaller than the 450\,$\mu$m beam.  There is thus not an apparent astrometric offset in our image.

\section{Source Detection and Number Counts}\label{sec_extract_counts}
\subsection{Source Extraction}\label{sec_extraction}

We first constructed a PSF for source extraction.  We generated a synthetic
PSF by averaging the 10 highest signal-to-noise ratio (S/N) sources in our final match-filtered map. These 10 
sources contribute equally to the averaged PSF.   The resulting PSF has a FWHM of $10\farcs1$, which is 9\%
broader than an idealized PSF (i.e., instrumental PSF with the same match-filtering), likely caused by the minor smearing effects from pointing
errors and focus changes.  In the steps described below, we tried using the synthetic PSF and the idealized PSF. The results are in agreement with each other,
except that the idealized PSF has a slight tendency to ``detect'' more faint sources very close to bright sources.   This is likely caused 
by its narrower profile and under-subtraction of the outer part of bright sources.  Since the synthetic PSF contains the smearing effect from observations
and should be more realistic, we adopt the synthetic PSF for the subsequent analyses.

To extract sources, we employ an iterative
procedure that is similar to ``CLEAN'' deconvolution in radio interferometric imaging \citep{hogbom74}.
We identify the peak pixel in the S/N map and subtract  5\% of a peak-scaled synthetic PSF at its position.
This 5\% subtraction fraction is often called CLEAN ``gain,'' and is typically set to a few to 10-percent in order to achieve stable convergence. 
We record its coordinates and subtracted flux.   We identify
the next peak in the image and repeat the process until we meet a floor S/N threshold, which we set to be 3.5 $\sigma$. 
During the process, if a subsequent S/N peak is located within $4\arcsec$ (approximately half the beam FWHM) from a previously identified peak, 
we consider them as the same source. In such a case, we only perform the iterative subtraction at the original position.  
Otherwise, we consider it as a new source and perform the PSF subtraction at the new position. 
Since the cleaning of source flux stops at 3.5 $\sigma$, we sum up the cleaned fluxes and the remaining 3.5 $\sigma$ flux to be the final flux for each individual source.
The very outer part of the map receives much shorter integration and contains arc and stripe patterns that are clearly not random noise.
Visual inspection shows that sources extracted in such areas are not always convincing.
We therefore only include sources extracted within $7\farcm5$ (r.m.s.\ noise $<10$ mJy) from the map center in our final catalog.
We detect 141 sources at 450\,$\mu$m at $>3.5$ $\sigma$, among which 98 are $>4.0~\sigma$ (circles in Fig.~\ref{fig1}).

The above iterative algorithm is inspired by the radio CLEAN deconvolution, and is similar to that in \citet{wang04}. 
It has the capability of separating mildly blended sources whose separation is larger than roughly one beam FWHM.  
For example, a pair of blended SCUBA sources in \citet{wang04} was subsequently confirmed by the interferometric 
imaging of \citet{wang11} to be multiple sources. In our 3.5 and 4-$\sigma$ catalogs, there are seven and five pairs, respectively, 
whose separations are less than $12\arcsec$ (see Fig.~\ref{fig1} for the 4-$\sigma$ cases).  They will be targets of our future followup observations, to test the 
accuracy of our source extraction in such limiting cases.  We also note that when the CLEAN gain is set to 1.0, our source extraction reduces to the
standard source identification adopted by other teams for SCUBA-2 images.  The few close pairs would have been identified as single sources
if the gain were 1.0.  In our simulations (Section~\ref{sec_a1}), this would increase the fraction of sources that are blends of multiple faint sources (hereafter the
``multiple fraction'') of the catalog.  
Also, in such a case, fluxes can be over-estimated for blended sources, since their fluxes are contaminated by the neighbors.  
However, the previous SCUBA-2 450\,$\mu$m surveys are shallower and further from the confusion limit.  Source blending and therefore the flux
contamination may not be an issue there.

\subsection{Number Counts and Simulations}\label{sec_counts}
\begin{figure}[t!]
%\epsscale{0.6}
\plotone{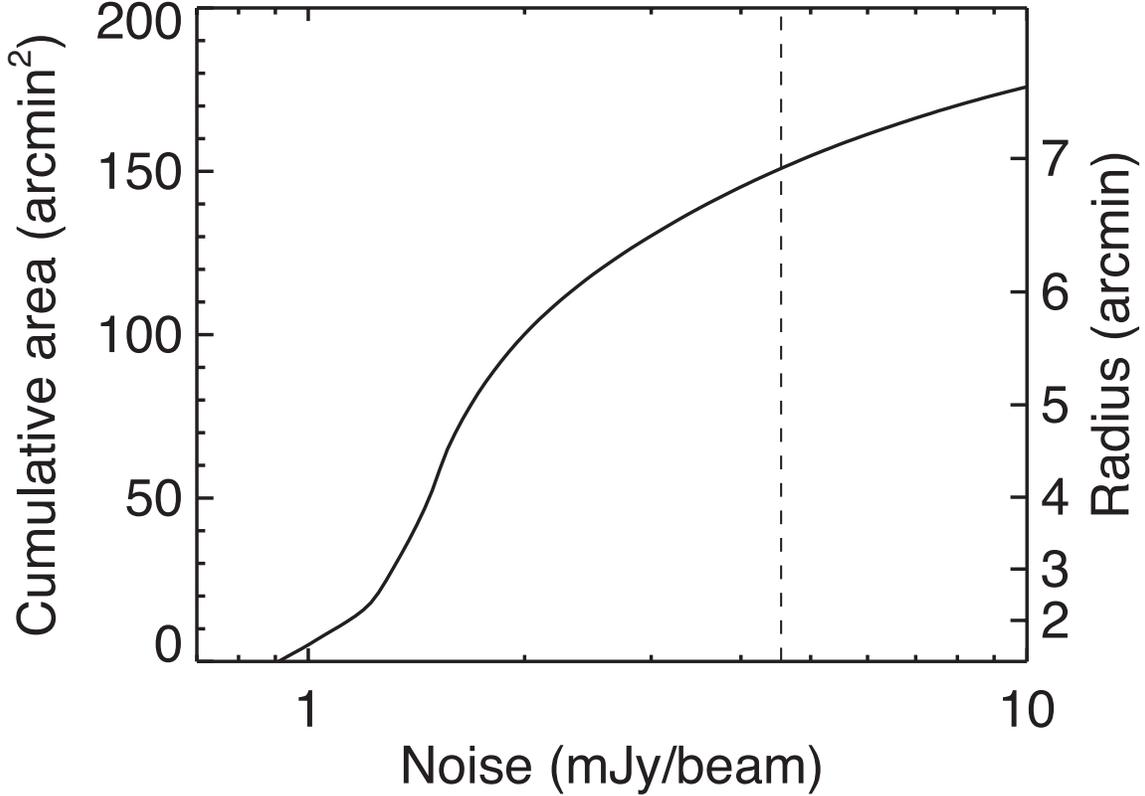}
\caption{Cumulative map area versus noise level.  The noise distribution in our image is roughly axisymmetric, 
so area can also be approximately mapped to radius, which is shown with the right-hand $y$-axis.
The map center has a noise of 0.91 mJy beam$^{-1}$.  The noise slowly increases
to $\sim2$ mJy beam$^{-1}$, where the map area reaches 100 arcmin$^2$. After that, the noise increases more rapidly toward
the outer region.  For our number counts, we do not use image area that is more than five times shallower
than at the map center, indicated by the vertical dashed line.
\label{fig3}} 
\end{figure}

We derived 450\,$\mu$m  source counts using our extracted sources and the noise map.
We only use 4~$\sigma$ sources for this calculation because they are more secure detections.  
We will discuss the reliability of the 4~$\sigma$ sources in Section~\ref{sec_a1}.
Furthermore, to ensure reliable source counts on faint sources, we confine our calculations to areas where the 
noise is less than five times the noise of the map center.  
This further discards the most noisy 9\% of the area used for the source extraction.
The cumulative area as a function of noise is shown in Fig.~\ref{fig3},
and the area involved in our estimates is 151 arcmin$^2$, indicated by the vertical dashed line.
97 of the 98 4-$\sigma$ sources fall in this area.  
%The outermost part of the area has a noise of 4.5 mJy, comparable to the deepest region in the survey of \citet{casey13}.

\begin{figure}[h!]
\epsscale{1.0}
\plotone{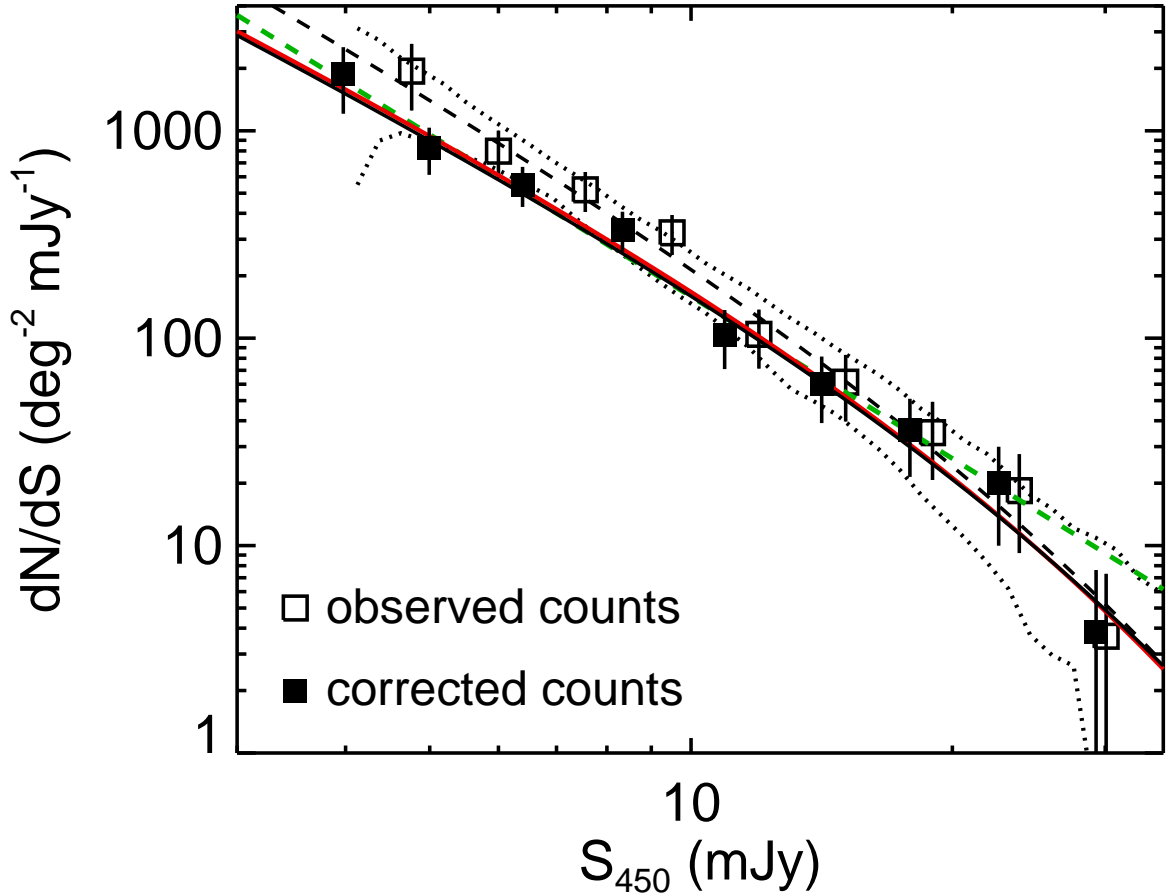}
\caption{Differential 450\,$\mu$m counts. Open squares are the observed raw counts and solid squares are the corrected counts.
The raw and corrected counts have different fluxes because of the correction for flux boosting (Section~\ref{sec_a1}).
The solid black curve is the input Schechter counts in the simulations.  The dashed black curve is the measured output counts in the simulations, and can
also be described by a Schechter function.  The two dotted black curves show the 68th-percentile range of the outputs of the 200 realizations.
The 68th-percentile range reasonably matches the observed error bars derived based on the assumption of Poisson errors.
Two parameterizations to the corrected counts are shown with colored curves.  
The red solid curve is a Schechter fit to the corrected counts.  The reduced $\chi^2$ of this fit is 0.45  ($N_{\rm df} = 6$).  This curve is almost indistinguishable 
from the input Schechter function (black solid curve), showing that our iterative procedure converges.
The green dashed line is a power-law fit to the corrected counts after excluding the brightest flux density bin.  The reduced $\chi^2$ of this fit is 0.38  ($N_{\rm df} = 6$).  
\label{fig4}} 
\end{figure}

In the number count calculation, each source contributes $1/(A_{\rm e}(S)dS)$ to the counts,
where $A_e$ is the effective area where the source can be detected at $>4~\sigma$ given its measured flux density $S$,
and $dS$ is the flux density interval.
Essentially, $A_{\rm e}(S)$ is the function in Fig.~\ref{fig3} when the $x$-axis is multiplied by 4.
The error is assumed to be Poissonian.  The faintest flux density bin is discarded, because it only contains one source. 
The resulting raw differential counts are presented in Fig.~\ref{fig4} (open squares)
and Table~\ref{tab_counts} (column 3).

%\begin{deluxetable}{cccccc}
%\tablewidth{0pt}
%\tablecaption{450\,$\mu$m Counts\label{tab_counts}}
%\tablehead{\colhead{$N$} & \colhead{$S_{\rm obs}$} & \colhead{Raw count}  &  & \colhead{$S_{\rm corr}$} & \colhead{Corrected count} \\
% \cline{2-3}  \cline{5-6}
% & \colhead{(mJy)} &  \colhead{(deg$^{-2}$ mJy$^{-1}$)} & & \colhead{(mJy)} & \colhead{(deg$^{-2}$ mJy$^{-1}$)}   }
%\startdata
%       9 	& 4.77	& $1938\pm686$	& & 3.94	& $1906\pm674$ 	\\
%      16  	& 6.00	& $797\pm204$	& & 5.02	& $814\pm209$ 	\\
%      21  	& 7.55	& $ 519\pm113$	& & 6.45	& $538\pm118$ 	\\
%      21  	& 9.51	& $322\pm70$		& & 8.36	& $331\pm72$ 		\\
%      10  	& 12.0	& $104\pm33$		& & 10.9	& $105\pm33$ 		\\
%       8   	& 15.1	& $61.4\pm21.7$	& & 14.1	& $61.3\pm21.7$ 	\\
%       6   	& 19.0	& $35.0\pm14.3$	& & 17.9	& $35.8\pm14.6$ 	\\
%       4   	& 23.9	& $18.4\pm9.2$	& & 22.7	& $19.6\pm9.8$ 	\\
%       1   	& 30.1 	& $3.65\pm3.65$	& & 29.0	& $3.97\pm3.97$ 	
%\enddata
%\tablecomments{$S_{\rm obs}$ is the observed flux density, and $S_{\rm corr}$ is the flux density corrected for boosting (Section~\ref{sec_a1})}.
%\end{deluxetable}

\begin{deluxetable}{ccccccccc}
\tablewidth{0pt}
\tablecaption{450\,$\mu$m Counts\label{tab_counts}}
\tablehead{\multicolumn{6}{c}{Differential counts} & & \multicolumn{2}{c}{Cumulative counts} \\
\cline{1-6} \cline{8-9}
\colhead{$N$} & \colhead{$S_{\rm obs}$} & \colhead{Raw $dN/dS$}  &  & \colhead{$S_{\rm corr}$} & \colhead{Corrected $dN/dS$}  &  & \colhead{$S_{\rm corr}$} & \colhead{Corrected $N(>S)$} \\
%\cline{2-3}  \cline{5-6}  \cline{8-9}
 & \colhead{(mJy)} &  \colhead{(deg$^{-2}$ mJy$^{-1}$)} & & \colhead{(mJy)} & \colhead{(deg$^{-2}$ mJy$^{-1}$)}    & & \colhead{(mJy)} & \colhead{(deg$^{-2}$)}  }
\startdata
       9 	& 4.77	& $1938\pm686$	& & 3.94	& $1906\pm674$ 	& & 3.57 	& $ 4662 \pm 709 $	\\
      16  	& 6.00	& $797\pm204$	& & 5.02	& $814\pm209$ 	& & 4.45 	& $ 3132 \pm 363 $	 \\
      21  	& 7.55	& $ 519\pm113$	& & 6.45	& $538\pm118$ 	& & 5.64  	& $ 2252 \pm 273 $	\\
      21  	& 9.51	& $322\pm70$		& & 8.36	& $331\pm72$ 		& & 7.29  	& $ 1419 \pm 202 $	\\
      10  	& 12.0	& $104\pm33$		& & 10.9	& $105\pm33$ 		& & 9.55  	& $ 730 \pm 136 $	\\
       8   	& 15.1	& $61.4\pm21.7$	& & 14.1	& $61.3\pm21.7$ 	& & 12.5 	& $ 447 \pm 103 $	 \\
       6   	& 19.0	& $35.0\pm14.3$	& & 17.9	& $35.8\pm14.6$ 	& & 15.9 	& $ 254 \pm 76.7 $	 \\
       4   	& 23.9	& $18.4\pm9.2$	& & 22.7	& $19.6\pm9.8$ 	& & 20.0 	& $ 119 \pm 53.2 $	 \\
       1   	& 30.1 	& $3.65\pm3.65$	& & 29.0	& $3.97\pm3.97$ 	& & 25.7 	& $ 22.8 \pm 22.8 $ 	\\
\enddata
\tablecomments{$S_{\rm obs}$ is the observed flux density. $S_{\rm corr}$ is the flux density corrected for boosting (Section~\ref{sec_a1}).  For the differential counts, flux densities are the
center of the bins.  For the cumulative counts, flux densities are the lower ends of the bins.}
\end{deluxetable}

The raw counts suffer from various observational biases: flux boosting 
produced by the Eddington bias and faint confusing sources; detection incompleteness; spurious sources;
and source blending. These are analyzed in detail in Section~\ref{sec_a1}.
To overcome these, we performed Monte Carlo simulations to derive the intrinsic counts.
We first created a ``true noise'' map using the jackknife method, described by \citet{cowie02} for SCUBA imaging.
We divided the half-hour SCUBA-2 scans into two interlacing halves and made two maps of nearly identical area
coverage and sensitivity.  The two half-maps underwent the same beam-matched convolution as the full map.
They were subtracted from one another, and each pixel of the resultant map was scaled 
by $\sqrt{t_1t_2}/(t_1+t_2)$ where $t_1$ and $t_2$ are the noise-weighted integration times of that pixel in the two half-maps.  
($t_1$ and $t_2$ are not identical because a fixed sky position can be swept by different bolometers in different scans.)
This effectively 
removes any faint sources and provides a map whose noise distribution is identical to the real image.  We measured
rms noise locally on the true noise map and found it consistent with the r.m.s.\ map generated by SMURF.
The brightness distribution in the true noise map is presented in Fig.~\ref{fig2} (red histogram).  It is symmetric about zero.
It is Gaussian-like, but not exactly a Gaussian.  Instead, it is a sum of Gaussians of various widths, because of the non-uniform sensitivity distribution,

We then created simulated images using the synthetic PSF. We randomly placed scaled PSFs in the true noise map 
with assumed source counts (see below) and flux densities between 1 and 50 mJy.  We found that this 1--50 mJy flux density range is sufficient, 
and the results do not change if we expand the range for the input sources (see Section~\ref{sec_mle}).
In each fine flux density bin, the number of the simulated sources is determined by the assumed counts plus a Poissonian fluctuation.
We created 100 simulated images using each of the positive and negative true noise maps.
Because all of the effects of flux boosting, completeness, spurious sources, and source blending depend on the 
intrinsic source counts, the selection of the intrinsic counts in the simulations may be crucial.
We employed an iterative procedure to approach the intrinsic counts from the observed raw counts.
We started with assuming that the intrinsic counts are the observed raw counts; we fit the raw counts with a Schechter function, 
and used the fitted function as input to create the first set of 200 realizations.  We then ran source extraction on the simulated images, 
derived simulated output counts, and compared them with the observed raw counts.  
We calculated the ratio between the simulated and observed counts, used that to adjust the input counts, and repeated the simulations.
In the first two iterations, the simulated output counts fluctuated around the observed counts, and then converged after the third iteration.
In the third iteration, the required adjustment was 
much smaller than the error bars in the counts.  Our final corrected counts ($C_{\rm corr}$) are calculated using the raw counts
($C_{\rm raw}$) and the ratios between the input and output counts in the simulations ($C_{\rm sim,in}$ and $C_{\rm sim,out}$, respectively):
\begin{equation}\label{eq_correction}
C_{\rm corr}(S_{\rm corr}) = C_{\rm raw}(S_{\rm obs}) \times \frac{C_{\rm sim,in}(S_{\rm corr})}{C_{\rm sim,out}(S_{\rm obs})},
\end{equation}
where $S_{\rm obs}$ is the observed flux density of sources and $S_{\rm corr}$ is the flux density corrected for boosting (see Section~\ref{sec_a1}).  
The corrected counts are presented in Table~\ref{tab_counts} (column 5)
and in Fig.~\ref{fig4} (solid squares).  For readers' reference, we also present the corrected cumulative counts in Table~\ref{tab_counts} (column 7), derived
with the same manner as the differential counts.

Our calculation in Eq.~\ref{eq_correction} is fundamentally different from what is often
adopted in the literature for mm/submm single-dish source counts, where the counts are corrected with completeness and spurious fractions.  
For such corrections, one generally has to assume one-to-one relations between the input  and output
source lists. Such an assumption is not accurate for single-dish observations, because multiple sources in the input list can be blended in the image and
detected as a single source of very different flux densities in the output list.  In contrast, Eq.~\ref{eq_correction} does not 
rely on such an assumption. %  and does not suffer from the complication caused by blended sources.  
The only part in Eq.~\ref{eq_correction} that contains a one-to-one relation is the correction of observed flux density ($S_{\rm obs}$) to intrinsic flux density ($S_{\rm corr}$), and
this is presented in Section~\ref{sec_a1}.  However, this only slightly affects the interpretation of where the counts are measured in terms of intrinsic flux density, 
and does not change the measured amplitude and slope of the counts.  % We will discuss this further in Section~\ref{sec_a1}.

The corrected counts in Fig.~\ref{fig4} can be reasonably well fit with a Schechter function:
\begin{equation}\label{eq_schechter}
\frac{dN}{dS} = \left(\frac{N_0}{S_0}\right) \exp\left(-\frac{S}{S_0}\right)\left(\frac{S}{S_0}\right)^{\alpha} \mathrm{~deg^{-2}~mJy^{-1}},
\end{equation}
with the best-fitting parameters listed in Table~\ref{tab_fit}. The reduced $\chi^2$ of the fit is 0.45, for a degree of freedom of $N_{\rm df} = 6$.  
The fitted function is the red solid curve in Fig.~\ref{fig4}. 
This fitting is considered as the final (fourth) iteration to approach the intrinsic counts.
It is almost indistinguishable from the input Schechter function used to create the simulations 
(black solid curve in Fig.~\ref{fig4}), showing that the iteration converges very nicely.
In the studies of \citet{chen13b} and \citet{hsu16}, similar iterative procedures were also adopted and the fitting of the intrinsic counts was made 
as part of the analysis procedure.

In the literature, mm/submm counts are also often fitted with a broken double-power law.
The turn-over point is typically between 20 and 30 mJy for 450\,$\mu$m counts \citep{casey13,chen13a,chen13b,hsu16,zavala17},
and 8.4 mJy for 850\,$\mu$m counts constrained by ALMA \citep{simpson15b}.  These two are consistent given that the 
$S_{850}/S_{450}$ ratio is typically 2.5 with large scatter \citep{casey13,hsu16}.
Our data seem to also suggest a turn-over point at 20--30 mJy, but we do not have sufficient data
to constrain the slope in the brighter portion.  If we fit our data with a broken power law, we get a
turn-over flux density between our first and second brightest flux density bin, and thus no constraint on the bright-end slope.
This is essentially fitting the counts with a single power law after excluding the brightest flux density bin.
We therefore fit our counts at $S<25$ mJy with a power-law form:
\begin{equation}\label{eq_powerlaw}
\frac{dN}{dS} = N_0 \left(\frac{S}{\rm mJy}\right)^{\alpha}\mathrm{~deg^{-2}~mJy^{-1}},
\end{equation}
with the best-fitting parameters listed in Table~\ref{tab_fit}. The reduced $\chi^2$ of the fit is 0.38 ($N_{\rm df} = 6$).  
This is the green dashed line in Fig.~\ref{fig4}
and is a slightly better fit to the data at $<25$ mJy, compared to the Schechter fit.  If we exclude the brightest flux density bin and then perform
a Schechter fit, we obtain a large $S_0$ ($>60$ mJy) and strong degeneracies among the fitted parameters.  This again shows that the 
data at $<25$ mJy are consistent with a single power law.

%\begin{deluxetable}{ccc}
%\tablewidth{0pt}
%\tablecaption{Parameterizations for the Corrected Counts\label{tab_fit}}
%\tablehead{\colhead{Parameter} &  \colhead{Schechter Fit}  &  \colhead{Power-Law Fit} \\
% & \colhead{(Eq.~\ref{eq_schechter})} & \colhead{(Eq.~\ref{eq_powerlaw})}   }
%\startdata
%$N_0$ 	& $2136\pm189$ 			& $62400\pm13400$ 	\\
%		& (deg$^{-2}$)				& (deg$^{-2}$~mJy$^{-1}$) \\
%\hline
%$\alpha$ 	& $-2.02\pm0.10$			& $-2.59\pm0.10$		\\
%\hline
%$S_0$ 	& $15.07\pm0.79$ 			&\nodata	\\
%		& (mJy)	&
%\enddata
%\end{deluxetable}

\begin{deluxetable}{ccc}
\tablewidth{0pt}
\tablecaption{Parameterizations for the Corrected Counts\label{tab_fit}}
\tablehead{\colhead{Parameter} &  \colhead{Schechter Fit}  &  \colhead{Power-Law Fit} \\
 & \colhead{(Eq.~\ref{eq_schechter})} & \colhead{(Eq.~\ref{eq_powerlaw})}   }
\startdata
$N_0$ 	& $2136\pm189$ deg$^{-2}$		& $62400\pm13400$ deg$^{-2}$~mJy$^{-1}$	\\
$\alpha$ 	& $-2.02\pm0.10$				& $-2.59\pm0.10$		\\
$S_0$ 	& $15.07\pm0.79$ mJy			&\nodata				\\
\hline
$\chi^2$	& 0.45						& 0.38				\\
$N_{\rm df}$	& 6						& 6					\\
\enddata
\tablecomments{The fitting is conducted on the bright-end counts above 3 mJy.  For the power-law fit, the
brightest flux-density bin is not included.}
\end{deluxetable}

\section{Fluctuation Analyses for the Faint End}\label{sec_mle}

\begin{figure}[t!]
%\epsscale{0.4}
\plotone{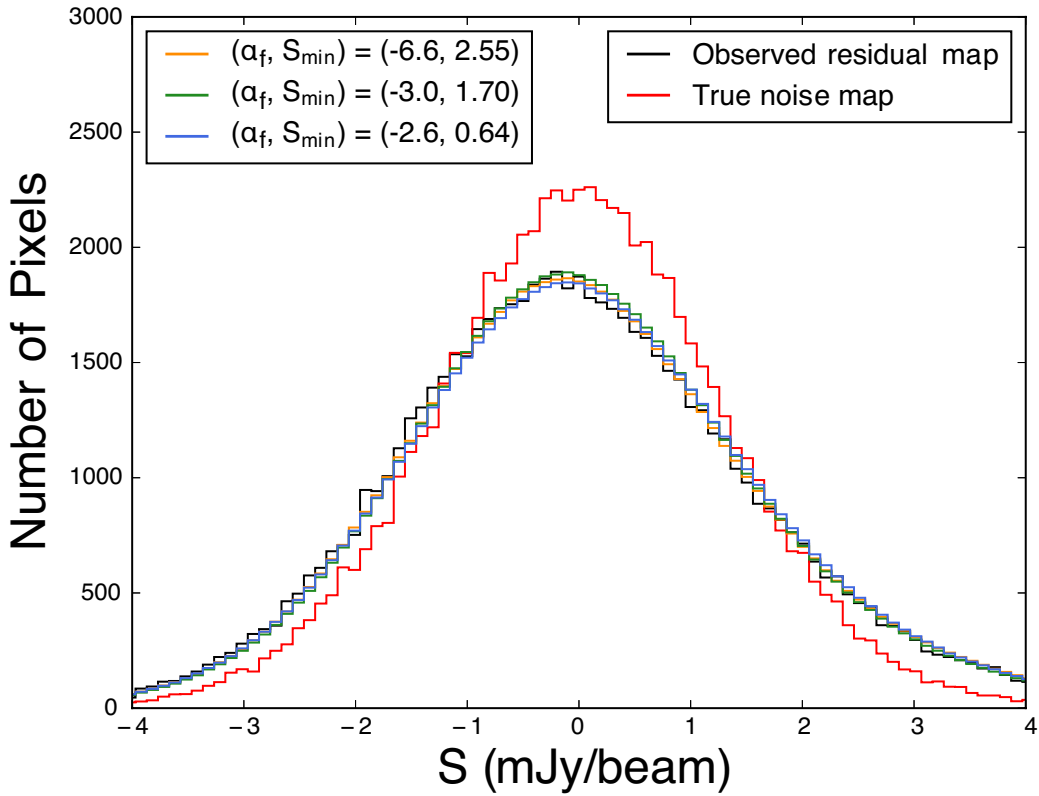}
\caption{Distributions of pixel brightness in the residual maps from the observations (black), the true noise map (red), and three representative models 
from our fluctuation analysis that have high likelihoods (orange, green, and blue, corresponding to the three crosses in Fig.~\ref{fig6}).
Here we only show the pixels within $5\arcmin$ of the map center, i.e., the pixels included in our fluctuation analysis.
\label{fig5}} 
\end{figure}

The derivation of the source counts in the previous section makes use of sources detected at 4 $\sigma$ at above 3.5 mJy (intrinsic).
The flux distribution of the residual map after 4-$\sigma$ sources are removed contains additional information about fainter sources. 
In Fig.~\ref{fig5}, we show the flux distribution of the pixels within $5\arcmin$ of the map center in the residual map
and in the true noise map.  The noise level in this area is 0.91--1.5 mJy.  Comparing to pure noise fluctuations, the residual map has 
excesses on both the positive and negative sides, caused by $<4~\sigma$ faint sources and the negative rings around their PSF.  This shape can 
be used to further constrain the faint-end counts.  To do this, we employ a parameter estimation method similar to that described in 
\citet[see discussion therein]{patanchon09}.  In the literature, \citet{maloney05}, \citet{coppin06}, \citet{weiss09}, and \citet{scott10} also used similar 
methods to derive counts from their mm/submm images.  The idea is to minimize the difference in the flux density distributions of the model-predicted and the 
measured residual maps, by maximizing the likelihood.  In the literature, this is often referred as the $P(D)$ method.

%For each number count model, the likelihood of the data is the probability that all of the flux values in the residual map had occurred:
%\begin{equation}\label{eq:def_likelihood}
%L(d|\theta)=\prod_{k}p(d_{k}|\theta),
%\end{equation}
For each number count model, the likelihood of the data is the probability that all of the flux density values in the residual map had occurred, in a logarithmic form:
\begin{equation}\label{eq:def_loglh}
\ln L(\theta) = \sum_{k}\ln (p(d_{k}|\theta)),
\end{equation}
where $\theta$ represents the generalized model parameters, $d=\{d_{1},...,d_{k},...\}$ is the set of flux density values of the pixels, 
and $p(d_{k}|\theta)$ is the probability distribution function (PDF) of  individual flux values with respect to the model.  
%Conventionally it is often expressed in a logarithmic form:
%\begin{equation}\label{eq:def_loglh}
%\ln L(\theta) = \sum_{k}\ln (p(d_{k}|\theta)).
%\end{equation}
Assuming that the PDF does not vary strongly in a flux density bin, then we have
\begin{equation}\label{eq:def_binloglh}
\ln L(\theta) = \sum_{i}n_{i}\ln (p_{i}(\theta)),
\end{equation}
where $n_{i}$ is the number of pixels in the $i$-th flux density bin, and $p_{i}(\theta)$ is the integral of the PDF in the $i$-th bin. %, i.e.,
%\begin{equation}\label{eq:def_binprob}
%\int_{d_{k}-\Delta_{1}}^{d_{k}+\Delta_{2}} p(d_{k}|\theta) d(d_{k}) = p_{i}(\theta),
%\end{equation}
%where $(d_{k}-\Delta_{1})$ and $(d_{k}+\Delta_{2})$ are two boundaries of $i$th bin. 
Eq.~\ref{eq:def_binloglh} is what we adopt for calculating the log-likelihood for a number count model, and the PDF is equivalently
the flux density histogram of the pixels in an image.

For simplicity, we adopt a single power law for the faint-end counts, with only two parameters.  The first one is the termination point of the counts, $S_{\rm min}$. 
The counts are assumed to be zero at $S<S_{\rm min}$ and $S_{\rm min}$ can be chosen to be much fainter than the detected sources.  
The second parameter is the faint-end power-law slope, denoted as $\alpha_{\rm f}$.   The power law is normalized 
to the corrected bright-end count at 3.94 mJy.  This is the median flux density in the faintest flux density bin of our
corrected bright-end counts.  In principle, we could add a third parameter so that the junction point of the bright and faint 
ends does not have to be 3.94 mJy.  Furthermore, we should allow a faint-end turn-over at perhaps below 1 mJy in order to prevent the sources
from over-producing the EBL, and this would require two additional parameters (the turn-over flux density and the slope in the extreme faint end).
However, as we will show, the results of the two-parameter model are already highly degenerate, and our data do not allow
meaningful constraints on additional parameters.

To determine the PDF of each number count model, we performed Monte Carlo simulations for the 450\,$\mu$m images in a way identical to that described 
in Section~\ref{sec_counts}.  Because the clustering of faint 450\,$\mu$m sources on the scales of our beam size is unknown (although likely to be weak),
the spatial distribution of our simulated sources is random (cf.\ \citealt{vernstrom14}, and see discussion therein).
For each faint-end count model, 200 simulated images were created using the true noise map, and another 200 using the negative true noise map.
The 400 images underwent the same source extraction procedure as the observed image, and 4-$\sigma$ sources were extracted and removed.
Since our goal is to constrain the faint-end counts, we only use the central $R<5\arcmin$ deep region, where the noise is less than approximately
twice the noise at the map center.  We also only used the flux density range of within $\pm4.2$ mJy for the likelihood calculation, so the results are not 
skewed by the small number of brighter pixels.  The resultant PDF averaged from the 400 model images was used as $p(\theta)$ in Eq.~\ref{eq:def_binloglh} 
for the calculation of the log-likelihood.

\begin{figure}[t!]
%\epsscale{0.4}
\plotone{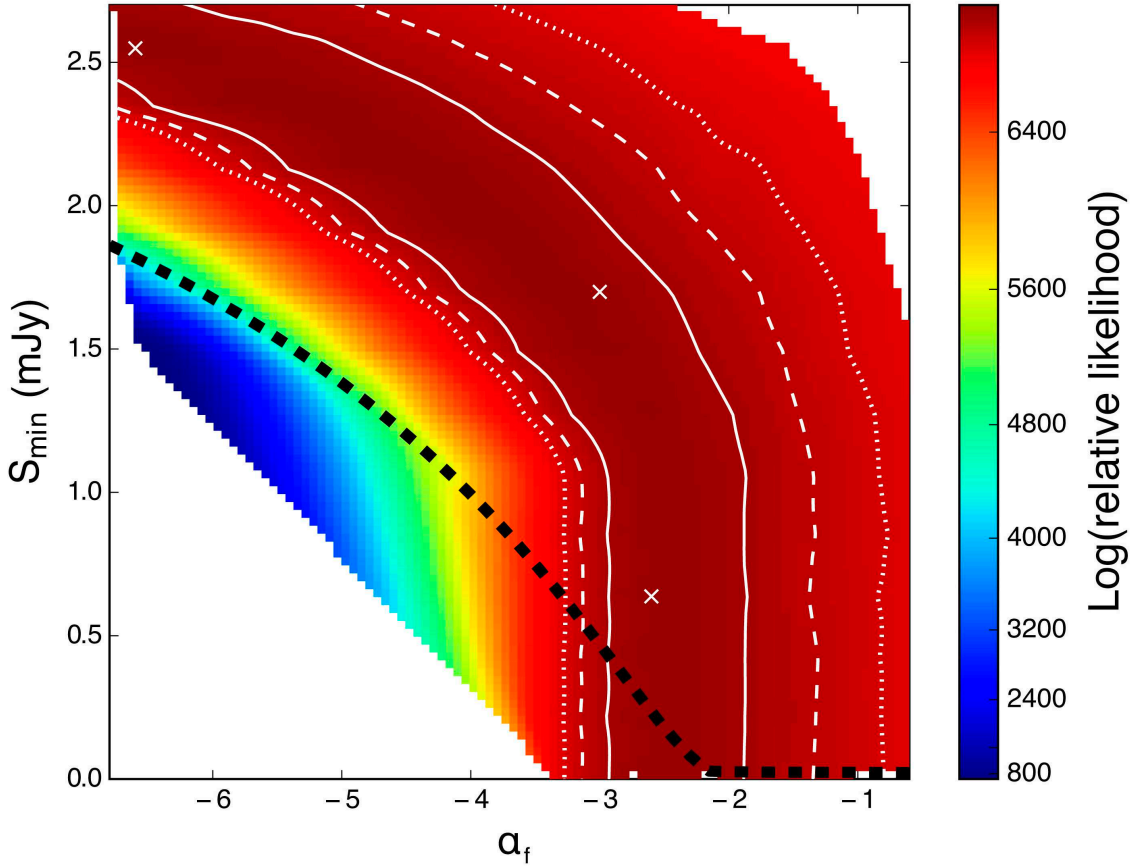}
\caption{Relative log-likelihood as a function of faint-end slope ($\alpha_{\rm f}$) and faint-end termination point ($S_{\rm min}$) in our fluctuation analysis. 
The white contours are 1-$\sigma$ (solid), 2-$\sigma$ (dashed), and 3-$\sigma$ (dotted) confidence ranges. The black dashed curve indicates where
the integrated surface brightness becomes 30\% higher than the \emph{COBE} EBL measured by \citet{gispert00}, in which 30\% is the typical uncertainty
in the \emph{COBE} measurements (see Section~\ref{sec_ebl} for more details).  
The cross ($\left[ \alpha_{\rm f}, S_{\rm min} \right]  = \left[-6.6, 2.55\right]$, $\left[-3.0, 1.70\right]$, and $\left[-2.6, 0.64\right]$) are picked from the 
high likelihood region and their PDFs are shown in Fig.~\ref{fig5}. 
\label{fig6}} 
\end{figure}

We calculated the log-likelihood over the parameter space of $\alpha_{\rm f} = -0.6$ to $-6.8$ and $S_{\rm min}$ from 0.0 to 2.75 mJy, with intervals of 0.2 in both.
We neglected areas in this parameter space if there we observed a trend of monotonically decreasing log-likelihood toward the extreme parameter values, or if 
the model overproduces the \emph{COBE} 450\,$\mu$m EBL measured by \citet{gispert00} by factors of more than four. The resultant relative 
log-likelihood distribution is presented in Fig.~\ref{fig6}.  There is a curved ``ridge'' of high likelihood extending from the upper-left to lower-right, representing the 
parameters that can reproduce the observed residual map.

To estimate the confidence levels, we used a bootstrap method.  Instead of using the PDF combined from the 400 simulated 
images for each model, we calculated the PDF and log-likelihood of each simulated image and then determined the dispersion in the log-likelihood.  
Because we perturbed the number of sources in each flux density bin according to the Poisson distribution in the simulations and adopted both the positive and
negative true noise maps, this dispersion represents the variance in the ensemble of images with the same set of model parameters and the underlying 
noise distribution.  It therefore offers an estimate of the confidence range for the likelihood.  This is the same method adopted by \citet{scott10}.
Around the maximum likelihood ``ridge''  in Fig.~\ref{fig6}, the mean dispersion is 61.  We therefore consider the 1, 2, and 3-$\sigma$ confidence 
ranges as 61, 122, and 183, respectively,
below the maximum log-likelihood.  These are the white contours in Fig.~\ref{fig6}.

To obtain a rough idea of how the above likelihood method can reproduce the observed residual map, we pick three sets of parameters 
(white crosses in Fig.~\ref{fig6}) that are well within the 1-$\sigma$ confidence range, and compare their PDFs with the observed one in Fig.~\ref{fig5}.  
We find that all three models closely reproduce the PDF of the observed residual map.  All of them are clearly different from pure noise.  We thus 
conclude that sources fainter than our 4-$\sigma$ detection limit are needed to explain the background fluctuation in the observed image.

\begin{figure*}[t!]
\epsscale{1.0}
\plotone{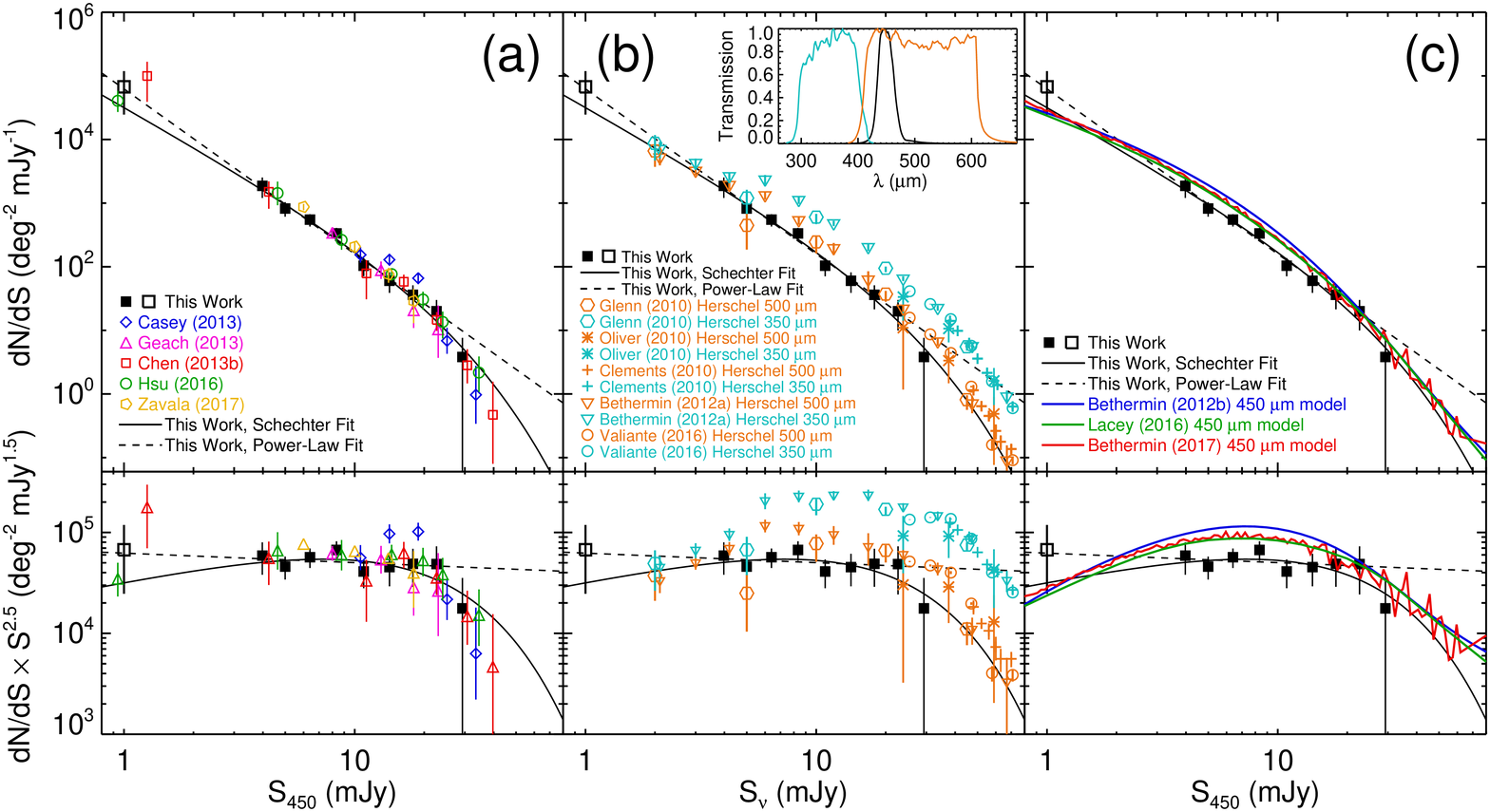}
\caption{Comparison of 450\,$\mu$m counts (top panels), and counts renormalized with $S^{2.5}$ to better show the differences and changes in shape (bottom panels).  
Panel (a) compares results from this paper and other SCUBA-2 450\,$\mu$m counts in the literature \citep{geach13,casey13,chen13b,hsu16,zavala17}.  
The solid squares are our counts derived from 4-$\sigma$ sources, while the open square is our 
count derived from a fluctuation analysis assuming that the faint-end counts extend to $\simeq1$ mJy (Section~\ref{sec_mle}).   
The counts derived from various SCUBA-2 surveys are in excellent agreement across the entire flux density range, but the data points
are not completely independent (see text).\\
Panel (b) compares our counts with \emph{Herschel} SPIRE 350\,$\mu$m and 500\,$\mu$m counts. The \emph{Herschel} counts were derived 
from direct source detection at the $\gtrsim20$ mJy bright end \citep{glenn10,clements10,oliver10,bethermin12a,valiante16}, and stacking analyses 
\citep{bethermin12a} and $P(D)$ analyses \citep{glenn10} at the faint end.
The $x$-axis shows the flux density measured at the corresponding wavebands, and we do not convert flux densities to a common wavelength.
The insert shows the filter profiles of the SPIRE 350\,$\mu$m (cyan) and 500\,$\mu$m (orange) wavebands, and the SCUBA-2 450\,$\mu$m wavebands (black).
Although it is reasonable to expect the 450\,$\mu$m counts to lie between the 500\,$\mu$m and 350\,$\mu$m counts, it is not the case here.  
At $>5$ mJy, nearly all \emph{Herschel} counts are consistently above the 450\,$\mu$m counts.  \citet{bethermin17} explained this with source
clustering at the scales of \emph{Herschel} beam sizes.\\
Panel (c) compares our counts with 450\,$\mu$m count models in \citet{bethermin12b,bethermin17} and \citet{lacey16}.
\label{fig7}} 
\end{figure*}

Unfortunately, both Fig.~\ref{fig6} and Fig.~\ref{fig5} show that a broad range of model parameters can meet the requirement of reproducing the observed 
low-level fluctuations. The 1-$\sigma$ confidence region in Fig.~\ref{fig6} is not localized.  Instead, it shows a strong degeneracy between 
$\alpha_{\rm f}$ and $S_{\rm min}$. We can reproduce the observations with a relatively shallower faint-end power-law slope $\alpha_{\rm f} > -3$ if the counts extend to 
$S_{\rm min} <1$ mJy.  We can also reproduce the observations with extremely steep power-law slopes ($\alpha_{\rm f} =-4$ to $-6$) if the counts terminate at 
$S_{\rm min}\simeq1.5$--2.5 mJy.  One can argue that the latter is highly unlikely for two reasons: (1) $\alpha_{\rm f} < -4$ is much steeper than the 
power-law slope we observed at the bright-end ($\alpha=-2.59\pm0.10$); and (2) observations in the lensing cluster fields clearly show that the nearly 
power-law counts should extend to at least 1 mJy (\citealp{chen13a,chen13b,hsu16}, also see Fig.~\ref{fig7} and Section~\ref{sec_count_compare}).  
Below we assume that the counts extend to $S_{\rm min}<1$ mJy, and discuss the implication of Fig.~\ref{fig6}.

At $S_{\rm min}<1$ mJy, the contours in Fig.~\ref{fig6}  become vertical.  This means that adding sources fainter than 1 mJy to the simulated maps does not improve
the results for a fixed power-law slope of $\alpha_{\rm f}= -2.6^{+0.4}_{-0.7}$.  In other words, our data are not sensitive to either a termination or a turn-over of the 
counts if it occurs at $<1$ mJy.  At 1 mJy, the 1-$\sigma$ range of the inferred counts is $2.5\times10^4$ to $1.2\times10^5$ deg$^{-2}$ mJy$^{-1}$ (Fig.~\ref{fig7}).   
This is consistent with the lensing counts in \citet{chen13b} and \citet{hsu16}, and also consistent with the power-law extrapolation of our bright-end counts.
These counts also translate to 0.3--1.4 sources per SCUBA-2 450\,$\mu$m beam, i.e., sources are highly confused. This is consistent with the confusion limit of
approximately 1 mJy estimated by \citet{geach13}, and explains why our fluctuation analysis is no longer sensitive to sources of $<1$ mJy. \citet{scott10} also 
found in their 1.1 mm analyses that adding sources fainter than the limit of roughly one source per beam does not alter the distribution.

To sum up, we do not find evidence of a turn-over of the counts between 1 and 3.5 mJy. A single power law with a slope of $\simeq-2.6$ can explain the
observations between 1 and 25 mJy. Our fluctuation analysis is insensitive to the counts at below 1 mJy as long as the counts maintain a power-law slope 
between $-1.9$ and $-3.1$.  A turn-over or a sudden termination of the counts at below 1 mJy can still be consistent with our data.  Indeed, the counts must 
turn over at some point below 1 mJy, or the integrated surface brightness from the 450\,$\mu$m sources will exceed the EBL. We will discuss this in Section~\ref{sec_ebl}.

\section{Discussion}\label{sec_discuss}

\subsection{Comparison with Previous Counts}\label{sec_count_compare}
We compare our counts with previous SCUBA-2 450\,$\mu$m counts in the literature in Fig.~\ref{fig7} (a).  
Among the blank-field surveys, the \citet{geach13} and \citet{zavala17} surveys reached r.m.s.\ sensitivities of $\lesssim2$ mJy
and 1.2 mJy, respectively, and their survey areas are comparable to ours (i.e., $\simeq150$~arcmin$^2$).  The \citet{casey13} survey covers a large area
of $\sim400$ arcmin$^2$ but with a shallower depth of 4.1 mJy.
The Geach et al. and Casey et al. surveys and ours are in the COSMOS field.  The Casey et al. survey area fully encloses ours, 
as well as that of Geach et al., while our survey area partially overlaps with that of Geach et al.  
The Zavala et al.\ survey field is in the Extended Groth Strip, a different line of sight.  
Both the surveys of \citet{chen13b} and \citet{hsu16} include lensing cluster fields of various depth and the COSMOS field
for their 450\,$\mu$m analyses. For COSMOS, both teams used the data taken by Casey et al.  The three cluster fields in 
Chen et al.\ with 450\,$\mu$m data are all included in Hsu et al., who added additional integrations, and also two further clusters.  
The sensitivities of the lensing-cluster surveys are determined by the amplification factors of the lensed sources, rather than the instrumental noise.
Because of the overlapping of data and survey fields, the results in these papers are not completely independent of each other, except for those in 
Zavala et al.

At the very bright end of $\gtrsim30$ mJy, all the SCUBA-2 counts are consistent with each other and show a steeper fall-off.  
Here, our count has a large error, because of our smaller effective area.  At the roughly 1\,mJy faint end, our result derived from the fluctuation
analysis and the extrapolation from our bright-end counts are both consistent with the lensing results of \citet{chen13b} and
\citet{hsu16}. As we mentioned in Section~\ref{sec_mle}, there is no evidence for a faint-end turn-over in our data at above 1 mJy.  
The two lensing data points at about 1 mJy seem consistent with this.

The counts are best constrained in our data below between $\gtrsim$ 3 mJy (the intrinsic flux density of our faintest detected source) and 25 mJy.  
Here, our results are consistent with previous ones within the error bars of each individual data point,
except for those between 10 and 20 mJy from the shallower survey of \citet{casey13}, which are higher than the other counts.
Cosmic variance (i.e., the effects of clustering on small fields) does not easily explain the elevated counts in Casey et al., 
since their survey field fully encloses ours and that of Geach et al. For cosmic variance to be the explanation, the $>10$\,mJy sources would have to be 
strongly clustered at $5\arcmin$ scales.  Unfortunately, there do not exist deep and wide-field 450\,$\mu$m dataset to test such clustering.  
A hint of clustering can be found in \citet{casey15} (also see \citealp{wang16}), who detected a filamentary structure at $z = 2.47$ in their 
SCUBA-2 image.  This structure, represented by their seven SCUBA-2 sources, runs through our field.  Five of their sources are within our survey area and
all of them are detected by us.  Therefore, this structure cannot be responsible for the difference between our counts and the counts in \citet{casey13}.  
\citet{hung16} reported another $z\sim2.1$ large-scale structure in this field. However, the only galaxy with a 450\,$\mu$m detection in their sample is
detected by both \citet{casey13} and us. So this structure does not seem to drive the higher counts in \citet{casey13}, either.

In order to more quantitatively compare the counts in these surveys, we refitted the counts quoted by the various authors and derived the counts at 6 mJy
and 10 mJy and the associated uncertainties.  The results are presented in Table~\ref{tab_count_compare}.  The dispersions in these values 
are 127 and 18 deg$^{-2}$ mJy$^{-1}$ at 6 mJy and 10 mJy, respectively, or 21\% and 10\% relative to the mean.
These are comparable to the shot noise in the counts, showing that the field-to-field variance should be even smaller.
We further tested this using the simulated 2 deg$^2$ 450\,$\mu$m catalog of \citet{bethermin17}, which contains source clustering, 
to conduct simulations of source extraction and number counts (see Section~\ref{sec_a2} for details).  
We found a scatter of about 10\% in the counts between 6 and 10 mJy for our field size. Therefore both the existing data and the simulations of \citet{bethermin17}
do not indicate a field-to-field variance that is much larger than 20\% for 450\,$\mu$m sources.
However, we caution that the scatter seen in Table~\ref{tab_count_compare} is not exactly the field-to-field variance, since the some of the fields overlap, 
and the Chen et al.\ and Hsu et al.\ results are combined from multiple fields.  
This might under-estimate the field-to-field variance, since counts from different fields are combined and averaged in these studies.
A more proper evaluation of the field-to-field variance will be provided by the S2CLS 450\,$\mu$m surveys (Geach et al., in prep.), along with
the few 450\,$\mu$m lensing fields.  Nevertheless, it is still quite remarkable that the various 450\,$\mu$m counts agree with each other at the roughly 20\% level.

\citet{geach17} reported a roughly 50\% field-to-field scatter for 850\,$\mu$m sources in S2CLS, whose field sizes are $0\arcdeg{\hskip -0.1cm}.5$ to $1\arcdeg$. 
The scatter is still roughly 50\% when the S2CLS 850\,$\mu$m counts are measured over $10\arcmin$ fields at 3.5 mJy, of which roughly 60\% is contributed
by the Poisson errors in such smaller fields (J.\ Geach, personal communication).  Both these results are much larger than the scatter we see in 450\,$\mu$m counts.
This suggests that the 450\,$\mu$m sources are less clustered at scales larger than our field size ($\gtrsim8$ Mpc at $z=2$).  
Two factors may contribute to this.  First, 850\,$\mu$m surveys are only sensitive to the most 
luminous dusty galaxies because of the confusion limit, while 450\,$\mu$m surveys can reach sources with lower luminosities (i.e., weaker clustering). However, 
the source densities listed in Table~\ref{tab_count_compare} are not yet at the limit of 850\,$\mu$m SCUBA-2 surveys.  Therefore, this is unlikely to be a luminosity 
effect.  Another reason is the strong negative $K$-correction at 850\,$\mu$m, making the 850\,$\mu$m band very sensitive to luminous dusty galaxies at $z>2$.  
\citet{wilkinson17} show that such high-redshift SMGs are clustered more strongly than those at lower redshifts. Such high-redshift SMGs are less abundant in 450\,$\mu$m images.
This is evident in the redshift distributions of the 450 and 850\,$\mu$m sources \citep{casey13}.
It is therefore possible that the larger field-to-field variance at 850\,$\mu$m is driven by the high-redshift SMGs.

\begin{deluxetable}{lcc}
\tablewidth{0pt}
\tablecaption{Comparison of 450\,$\mu$m Counts\label{tab_count_compare}}
\tablehead{\colhead{Authors} &  \colhead{$dN/dS$~(6 mJy)}  &  \colhead{$dN/dS$~(10 mJy)} \\
 & \colhead{(deg$^{-2}$~mJy$^{-1})$} & \colhead{(deg$^{-2}$~mJy$^{-1}$)}   }
\startdata
This work\tablenotemark{a}			&  $602.2 \pm 66.8$		& $160.4 \pm 18.6$	\\
\citet{geach13}\tablenotemark{b}		&  $747.0 \pm 165.0$	& $180.3 \pm 18.4$	\\
\citet{casey13}\tablenotemark{b} 		& \nodata 				& $180.5 \pm 28.4$	\\
\citet{chen13b}\tablenotemark{a} 		&  $530.7 \pm 172.0$ 	& $145.7 \pm 30.5$	\\
\citet{hsu16}\tablenotemark{a} 			&  $563.7 \pm 85.5$		& $170.6 \pm 24.7$	\\
\citet{zavala17}\tablenotemark{b}		&  $827.9 \pm 64.4$		& $196.9 \pm 19.5$ \\
\enddata
\tablenotetext{a}{Power-law fits are adopted.}
\tablenotetext{b}{Schechter fits are adopted.}
\end{deluxetable}

In Fig.~\ref{fig7} (b), we show a comparison between our 450\,$\mu$m counts and \emph{Herschel} SPIRE 350/500\,$\mu$m counts, along
with the filter profiles of these three passbands. Given the passbands, is reasonable to expect the 450\,$\mu$m counts to fall between the 
350 and 500\,$\mu$m ones, probably closer to the 500\,$\mu$m ones. To our surprise, this is not the case, and the 450\,$\mu$m counts fall 
below both the \emph{Herschel} 350 and 500\,$\mu$m counts.  
The \emph{Herschel} confusion limit is approximately 20 mJy in both wavebands.  The counts at $>20$~mJy were derived with direct 
source extraction \citep{glenn10,clements10,oliver10,bethermin12a,valiante16}, while the counts at $<20$~mJy were derived by stacking 
\emph{Spitzer} 24 $\mu$m sources in the \emph{Herschel} images \citep{bethermin12a} and from $P(D)$ analyses \citep{glenn10}.  
The \emph{Herschel} counts derived by various authors are highly consistent with each other, regardless of how they were derived.
All of them lie significantly above the SCUBA-2 450\,$\mu$m counts over nearly the entire flux density range of interest. The 500\,$\mu$m data 
points are individually about 1 to 2 $\sigma$ away from the SCUBA-2 counts; however, the fact that collectively nearly all of them are above the 
SCUBA-2 points means that the over-abundance in the \emph{Herschel} counts is significant. In the literature, 
over-abundant \emph{Herschel} counts were already noted by \citet{chen13b}, but \citet{geach13} and \citet{zavala17} considered the 
\emph{Herschel} counts to be consistent with their SCUBA-2 counts, perhaps because of the larger error bars and sparse data points. With our 
better constrained SCUBA-2 450\,$\mu$m counts, it should now be clear that there is a discrepancy with the \emph{Herschel}-derived counts. 

\citet{bethermin17} offered an explanation for the elevated \emph{Herschel} counts.  With simulations of dark matter halos in a 2 deg$^2$ lightcone, 
an abundance-matching technique to populate the halos with galaxies, and models of star formation and galaxy SEDs, these authors can 
reproduce the observed \emph{Herschel} counts with much lower intrinsic counts. They therefore attribute the elevated \emph{Herschel} counts to source 
blending in the \emph{Herschel} images, caused by the clustering of FIR sources and the large \emph{Herschel} beam in the SPIRE bands.  
We note that just blending alone does not necessarily bias the $P(D)$ analyses and stacking analyses in the faint end, as the associated
simulations would have accounted for its effects. However, the simulations typically do not include source clustering, which can amplify the
observed image fluctuation and the stacking signal.

\citet{bethermin17} also show that the intrinsic counts are much closer to the SCUBA-2 counts, because the SCUBA-2 450\,$\mu$m beam is much smaller. 
Despite this, in Fig.~\ref{fig7} (c), we show that the 450\,$\mu$m counts predicted by \citet{bethermin17}, as well as by \citet{lacey16},
are somewhat higher than the SCUBA-2 counts at flux densities above 3 mJy.  The over-prediction in the counts is up to about 70\% in the \citet{bethermin17} case.
The slopes of the counts at flux densities below 3 mJy is also shallower in the models than what
the observations suggest.  The above 70\% offset in counts can be translated to a 23\% offset in flux, which is much larger than the generally expected
12\% calibration uncertainty in SCUBA-2 observations \citep{dempsey13}.  In addition, the offset between the model counts and observed counts
is not a constant.  It is not possible to explain this offset with a simple calibration error.  The discrepancy between the models and observations
thus appears to be real.
Nevertheless, the work of \citet{bethermin17} and the comparisons in our Fig.~\ref{fig7} (b) highlight the importance of high angular resolution in
studying high-redshift galaxies.  The beam size of SCUBA-2 at 450\,$\mu$m is much less affected by clustering (Section~\ref{sec_a2}), and can
therefore provide much more robust measurements 
of both the spatial density and flux density of sources.

Finally, we estimate the bias in the \emph{Herschel} 500\,$\mu$m counts, assuming that our 450\,$\mu$m counts are unbiased.
At around 10~mJy, the observed \emph{Herschel} 500\,$\mu$m counts are higher than our 450\,$\mu$m counts by 1.25 times in flux 
density, or 1.8 times in number density.  These are lower limits for the bias in the \emph{Herschel} counts, since 500\,$\mu$m counts 
should be intrinsically lower than 450\,$\mu$m counts.  Estimating the intrinsic offset between the 450 and 500\,$\mu$m counts requires
knowledge about the $S_{450}/S_{500}$ flux density ratios of the sources, which are determined by their redshift distribution and SEDs. 
A zeroth-order estimate can be made through the model of \citet{bethermin17}.  Although Fig.~\ref{fig7} (c) shows that the model does not
perfectly match the 450\,$\mu$m counts, it nevertheless predicts 450\,$\mu$m counts that lie between the 350 and 500\,$\mu$m counts,
meaning that the $S_{450}/S_{500}$ flux density ratios in its source catalog are not too far off.  If we use the intrinsic offset between the 450 and 
500\,$\mu$m counts in the model of \citet{bethermin17} to adjust the above observed offset, we then find that the observed 
\emph{Herschel} 500\,$\mu$m counts should be higher than the intrinsic 500\,$\mu$m counts by roughly 1.4 times in flux, or 2.5 times 
in number density, at flux densities of about 10 mJy.

\subsection{Contributions to the 450\,$\mu$m EBL}\label{sec_ebl}

We compare the 450\,$\mu$m EBL resolved in our SCUBA-2 observations with various \emph{COBE} EBL measurements.
The low angular resolution satellite EBL measurements and resolved source counts contain different sets of systematics.
Comparing them provides insight into the nature of the sources that give rise to the EBL, as well as testing them against each other.

The 450\,$\mu$m EBL estimations based on the \emph{COBE} FIRAS experiment are 109 Jy deg$^{-2}$ \citep{puget96}, 142 Jy deg$^{-2}$
\citep{fixsen98}, and 150 Jy deg$^{-2}$ \citep{gispert00}. The uncertainty in them is about 30\%. 
The difference between these results arises from the subtractions of the foregrounds, and should be considered as a measure of systematic uncertainty.
%\emph{Planck} data were also used by various teams to derive the EBL \citep{planckxxx,planckviii,lenz17}.
%We interpolated the estimates from the \emph{Planck} bands between 217 and 857 GHz to determine the \emph{Planck} 450\,$\mu$m EBL.
%The results are 190 Jy deg$^{-2}$ from Planck Collaboration XXX, 150 Jy deg$^{-2}$ from Planck Collaboration VIII, and 180 Jy deg$^{-2}$ from Lenz et al.
%The statistical uncertainties in these results are typically a few percent, while systematic uncertainties can be up to 20\%.
%The first two \emph{Planck} results adopted the \emph{Herschel}-based galaxy evolution model of \citet{bethermin12b} for the EBL and for galaxy shot noise.
%The overestimation in the model discussed in the previous subsection would propagate to the EBL measurements (e.g., overestimating the galaxy shot noise). However, this is not the case in the
%calculations of Lenz et al., who used \ion{H}{1} data to separate the Galactic dust from the EBL.  Nevertheless, the difference in these
%\emph{Planck} results should be considered as the systematic uncertainty, as for the \emph{COBE} results.
In the subsequent discussion, we compare our resolved EBL with the full range of 109--150 Jy deg$^{-2}$ from the above \emph{COBE} results.

\begin{figure}[t!]
%\epsscale{0.4}
\plotone{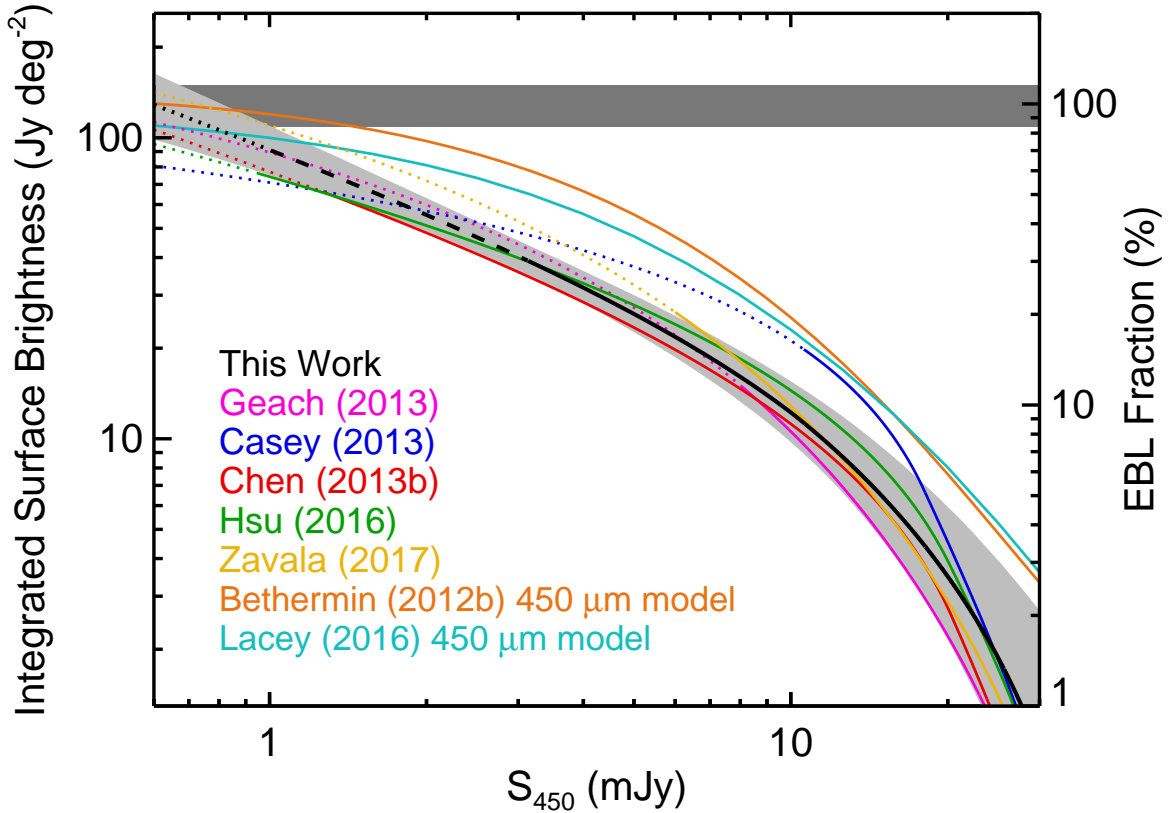}
\caption{Integrated 450\,$\mu$m surface brightness from various SCUBA-2 surveys. The dark shaded area shows the range of 450\,$\mu$m 
EBL values inferred from various \emph{COBE} studies. The black curve is the result derived from our 450\,$\mu$m counts, and
the light shaded area is its uncertainty range. The colored curves are results from previous 450\,$\mu$m surveys. 
The solid portions of the curves represent counts constrained by detected sources, 
while the dotted portions represent the faint-end extrapolations of the counts. The dashed portion of the black
curve is where our counts are constrained through the fluctuation analysis. The models of \citet{bethermin12b} and \citet{lacey16} are 
also shown for comparison.  The right-hand $y$-axis is the fraction of the resolved EBL, and the 100\% resolution level is set to be the mid point between the minimum
and maximum values among the \emph{COBE} determinations.
\label{fig8}} 
\end{figure}

We show the 450\,$\mu$m surface brightness integrated from our source counts in Fig.~\ref{fig8}. For this calculation, we adopted
our Schechter fit for above 10 mJy and our power-law fit for 0.1--10 mJy. For uncertainty estimation, we perturbed 
our counts according to their errors, re-conducted the fitting, and calculated the dispersion in the results.
Down to 3.5 mJy, which is the corrected flux density of our faintest 4-$\sigma$ 
source, we have resolved $35.5\pm4.3$ Jy deg$^{-2}$ of the 450\,$\mu$m EBL (black solid curve in Fig.~\ref{fig8}). This corresponds to $24\pm3\%$ to $33\pm4\%$ of the 
total EBL, depending on which EBL estimate we adopt. If we integrate down to 1 mJy, where we have constraints on the counts from our
fluctuation analysis, we obtain a value of $90.0\pm17.2$ Jy deg$^{-2}$ (black dashed curve in Fig.~\ref{fig8}), corresponding to $60\pm11\%$ to $83^{+15}_{-16}\%$ of the EBL.

If we continue the integration with the same power-law slope, we will reach 100\% resolution of the EBL from \citet{puget96} at 0.77 mJy, while 
in order to fully account for EBL from \citet{gispert00}, the counts have to extend to approximately 0.48 mJy with the same faint-end slope.
Roughly speaking, the flux of 0.48 mJy corresponds to infrared luminosities of $3.8\times10^{10}$~L$_\sun$
and $7.9\times10^{10}$~L$_\sun$ at $z=1$ and $z=2$, respectively, if we assume the luminosity-dependent dust SEDs of Chu et al.\ (in prep.),
which incorporate the latest \emph{WISE} and \emph{Herschel} photometry for local infrared selected galaxies \citep{chu17}.
The luminosity would be 30\% ($z=2$) to 40\% ($z=1$) higher if we assume the median SED of bright 870\,$\mu$m-selected 
SMGs in \citet{danielson17}, because such luminous SMGs have higher dust temperature and therefore more emission from
$\lesssim200$\,$\mu$m.  The luminosity of 3.8--$7.9\times10^{10}$~L$_\sun$ corresponds to SFR of roughly 6--13 M$_\sun$ yr$^{-1}$.  
These are in the regime of the typical SFR of rest-frame UV selected high-redshift galaxies.

The above assumes that the extension of the counts below 1 mJy maintains a power-law slope of $\alpha_f=-2.6$. A more likely scenario is that the
counts become shallower below 1 mJy.  Shallower (Schechter) slopes in the UV luminosity functions are typical for faint rest-frame UV selected populations, ranging from 
$-0.8$ to $-1.7$ at $z\simeq1.7$ to 3 \citep{steidel99,sawicki06,reddy09}. Steeper slopes are observed at higher redshifts: $-1.6$ at $z\simeq4$ \citep{bouwens15}, and $-2.0$ at $z\simeq6$--8 \citep{bouwens15,bouwens17}, but are still shallower than the slope we observed on 450\,$\mu$m sources.

Another hint of a shallower slope in the extreme faint end comes from ALMA imaging at longer wavelengths.
Recent ALMA observations reported nearly full resolution of the 1.1\,mm EBL with serendipitous detections of faint sources \citep{fujimoto16} and
stacking analyses \citep{aravena16,wang16b,dunlop17}.  Roughly speaking, it requires the detections of $S_{1100} \simeq0.01$--0.02 mJy sources
to achieve a full resolution of the 1.1 mm EBL.  For $z\simeq1$--2, these sources would have 450\,$\mu$m fluxes of about 0.1--0.2 mJy, a few times 
fainter than the above 0.48 mJy extrapolated based on a constant slope. This again suggests a slope shallower than $-2.6$.  

Our STUDIES survey was allocated 330 hrs of observations (including some overheads). Roughly 40\% of 
the observations were carried out in the first year and included in this paper. Once the survey is complete, we can expect to detect sources
with intrinsic flux densities of $\gtrsim 2$ mJy in the deepest region and directly resolve around 55 Jy deg$^{-2}$ of the EBL (30\% to 50\%).
Furthermore, an extension of STUDIES was recently approved to conduct an equally deep pointing in the Subaru/XMM-Newton Deep Survey
field \citep{furusawa08}.  Both the increases in depth and survey area will lead to dramatically improved source counts in the 2--30 mJy bright end.
The deeper image may enable meaningful fluctuation analysis below 2 mJy with more model parameters than our current simple two-parameter model.
If this is the case, better constraints in the 1--2 mJy regime may help to narrow down the parameter space for the extremely faint end below 1 mJy
(e.g., a shallower slope and/or a turn-over flux density).  It will be interesting if the improved fluctuation analysis can reach $>100\%$ resolution of the \emph{COBE} EBL,
and hence start to pin down the EBL better than the satellite estimates.  The most robust EBL estimate will likely
require extremely deep and wide interferometric imaging in the future.

\section{Summary}\label{sec_summary}

In our JCMT Large Program, STUDIES, we are carrying out extremely deep 450\,$\mu$m imaging with SCUBA-2 in the COSMOS-CANDELS region.
The $7\arcsec$ resolution of the 450\,$\mu$m band provides the opportunity to probe beyond the conventional 850\,$\mu$m confusion limit to
detect fainter galaxies.
Our goal is to detect faint 450\,$\mu$m sources close to the confusion limit of SCUBA-2, to study a representative sample of the high-redshift 
FIR galaxy populations that give rise to the bulk of the FIR background and the cosmic star formation.
With the first year of STUDIES data, we reached a noise of 0.91 mJy at the map center for point sources, about 30\% deeper than previous deepest 450\,$\mu$m map,
and covered a deep area of $R=7\farcm5$. 
We detected 98 and 141 sources at 4.0 and 3.5 $\sigma$, respectively.  Our source counts are best constrained
between 3.5 and 25 mJy (4 $\sigma$), and are consistent with most of the previous counts derived from blank-field and lensing-cluster surveys.  The field-to-field 
variance among the various surveys is about 20\%, much smaller than the variance observed in 850\,$\mu$m surveys, perhaps suggesting
weaker clustering in the 450\,$\mu$m population at scales larger than our field size.  In this flux density range, our counts are consistent with a power law with a slope of $\alpha=-2.59\pm0.10$.  
We further constrain the counts at 1--3.5 mJy with a fluctuation analysis.  We see evidence of a termination or turn-over of the faint-end
counts between 1 and 3.5~mJy.  The power-law slope at 1--3.5~mJy is $\alpha_{\rm f} = -2.6^{+0.4}_{-0.7}$, consistent with the slope at $>3.5$ mJy. This is also consistent 
with the counts at around 1 mJy derived from previous lensing-cluster surveys. On the other hand, ours and all other SCUBA-2 450\,$\mu$m
counts appear significantly lower than \emph{Herschel} counts at 350 and 500\,$\mu$m.  This discrepancy is likely caused by source blending under 
the coarse  \emph{Herschel} beam, amplified by source clustering at the scales of the beam.  Because of its higher angular resolution, SCUBA-2 
counts at 450\,$\mu$m do not suffer from these effects.  Our extremely deep 
SCUBA-2 map has resolved a substantial fraction of the 450\,$\mu$m EBL estimated by \emph{COBE}, 
$24\pm3\%$ to $33\pm4\%$ from the 4-$\sigma$ sources at $>3.5$ mJy, 
and $60\pm11\%$ to $83^{+15}_{-16}\%$ if we include the 1--3.5 mJy faint-end counts derived from the fluctuation analysis. 
STUDIES is an ongoing survey and we expect that our future deeper image can be used to better determine the EBL at 450\,$\mu$m, as 
well as providing accurate counts for constraining galaxy evolution models.

\acknowledgments

We thank the JCMT/EAO staff for the observational support and the data/survey management, 
Oliver Dor\'{e} and Matthieu B\'{e}thermin for a discussion on the EBL and galaxy evolution models,
Li-Yen Hsu for a discussion on the lensed source counts, David Sanders and Jason Chu for providing their local galaxy SED templates,
and the anonymous referee for the comments that significantly improve the manuscript.
W.H.W., W.C.L., C.F.L., and Y.Y.C. acknowledge support from the Ministry of Science and Technology of Taiwan grant 105-2112-M-001-029-MY3.
I.R.S. acknowledges supports from STFC (ST/P000541/1), the ERC Advanced Investigator programme DUSTYGAL (321334),
and a Royal Society/Wolfson Merit Award.  
X.Z.Z. acknowledges support from National Key R\&D  Program of China (2017YFA0402703) and NSFC (grant 11773076).
J.M.S. acknowledges support from the EACOA Fellowship and C.C.C. acknowledges support from the ESO Fellowship. 
W.I.C. acknowledges support from the European Research Council through the award of the Consolidator Grant ID 681627-BUILDUP.
H.D. acknowledges support from the Spanish Ministry of Economy and Competitiveness (MINECO) under the 2014 Ram\'{o}n y Cajal program MINECO RYC-2014-15686.
M.J.M acknowledges supports from the National Science Centre, Poland through the POLONEZ grant 2015/19/P/ST9/04010 and 
the European Union's Horizon 2020 research and innovation programme under the Marie Sk{\l}odowska-Curie grant agreement No. 665778. 
R.J.I. and I.O. acknowledge support from the European Research Council in the form of the Advanced Investigator Programme, 321302, {\sc cosmicism}.
X.W.S. acknowledges support from the Chinese NSF through grant 11573001.
J.L.W. acknowledges support from a European Union COFUND/Durham Junior Research Fellowship under EU grant agreement number 609412, and 
additional support from STFC (ST/P000541/1). SC, MS, and DS acknowledge support from the Natural Sciences and Engineering Research Council (NSERC) of Canada.

The James Clerk Maxwell Telescope is operated by the East Asian Observatory on behalf of 
The National Astronomical Observatory of Japan, Academia Sinica Institute of Astronomy and Astrophysics, 
the Korea Astronomy and Space Science Institute, the National Astronomical Observatories of China and the 
Chinese Academy of Sciences (Grant No. XDB09000000), with additional funding support from the Science and Technology 
Facilities Council of the United Kingdom and participating universities in the United Kingdom and Canada.

The authors wish to recognize and acknowledge the very significant cultural role and reverence that the summit of Maunakea 
has always had within the indigenous Hawaiian community.  We are most fortunate to have the opportunity to conduct 
observations from this mountain. 

The authors would like to dedicate this paper to the memory of Fred Kwok-Yung Lo, a JCMT Fellow in 1991, and a pioneer
of millimeter and submillimeter astronomy.

\appendix

\section{Flux Boosting, Completeness, Spurious Fraction, and Source Blending}\label{sec_a1}
The procedure for deriving true counts (Section~\ref{sec_counts}) treats the observational effects of flux boosting,  
completeness, spurious sources, and blending simultaneously.  This is different from some other works in the literature
\citep[e.g.,][]{casey13,geach13,chen13a,zavala17}.
Previous 450\,$\mu$m studies did not deal with the effects of source blending, partially because they
were relatively shallow and the probability of chance projections was lower.
The smaller 450\,$\mu$m beam also makes it less likely for two sources to blend with each other
\citep[e.g.,][]{cowley15}, unlike the case for longer wavebands \citep{wang11,karim13,hodge13,simpson15b}.   %smolcic12
Later we will show that source blending is not entirely negligible in our deep image.
For the other effects, previous studies estimated the amplitude of the corrections through simulations similar to ours,
and applied these corrections to the observed flux densities and the counts.   The correction to the counts typically looks like
\begin{equation}\label{eq_correction_old}
C_{\rm corr}(S_{\rm corr})= C_{\rm raw}(S_{\rm obs}) \times \frac{(1-\mathcal{F}_{\rm spu}(S_{\rm obs}))}{\mathcal{F}_{\rm comp}(S_{\rm corr})},
\end{equation}
where $\mathcal{F}_{\rm spu}$ is the fraction of spurious sources, and $\mathcal{F}_{\rm comp}$ 
is the detection completeness fraction.  
To derive the two $\mathcal{F}$ terms and the conversion between $S_{\rm obs}$ and $S_{\rm corr}$, 
one normally has to assume a one-to-one relation between an output source 
and an input source in the simulation.
The possibility of source blending makes this assumption no longer valid.
We avoid this and do not attempt to correct the observed counts using Eq.~\ref{eq_correction_old}.
Instead, we fully rely on the iterative procedure and Eq.~\ref{eq_correction} described in Section~\ref{sec_counts} to derive the intrinsic counts.
However, we can still estimate these effects using our simulations, in order to obtain a picture of our observation
and source extraction efficiency, and to compare with previous studies.

\begin{figure*}[t!]
\epsscale{1.0}
\plotone{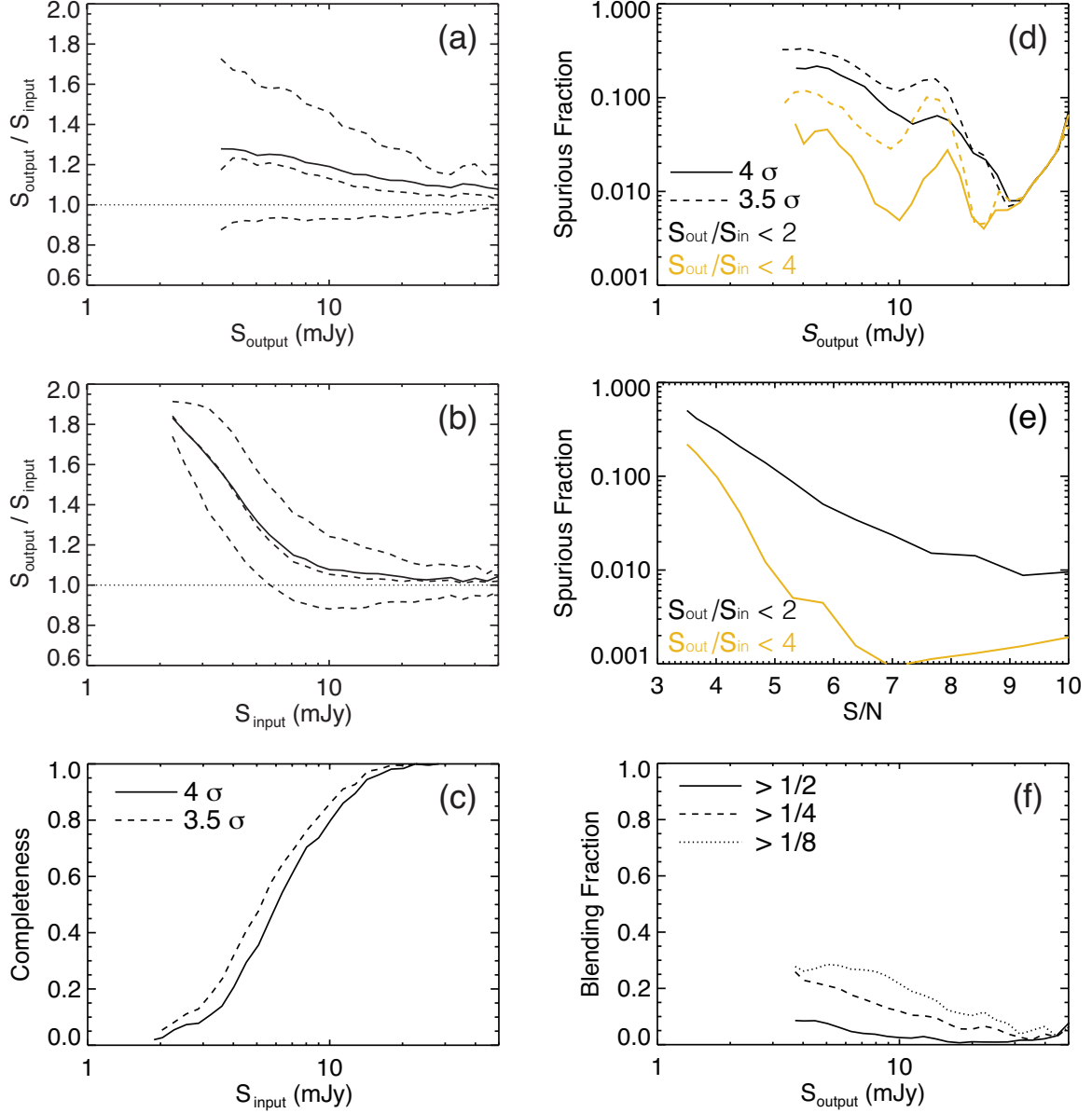}
\caption{Bias effects in the observations and source extraction. Panel (a) shows the flux boosting factor versus output (observed) source flux density. 
Panel (b) shows the flux boosting factor versus input (intrinsic) source flux density.  In both (a) and (b), the solid curves are mean values, while the 
dashed curves are the 68th ($\pm1$~$\sigma$) and 50th-percentile (median) values.  The overall boosting correction for measured flux densities
is around 20\%. A small number of intrinsically faint sources ($S_{\rm input} \gtrsim2$ mJy) can be strongly boosted into our detection range.
Panel (c) shows the completeness versus input (intrinsic) source flux density for a 4-$\sigma$ detection threshold (solid curve) and a 3.5-$\sigma$ 
detection threshold  (dashed curve).  Panel (d) shows the spurious fraction versus output (measured) flux density.  The spurious fraction for 4-$\sigma$ 
detections are shown with the solid curve, and for 3.5-$\sigma$ detections with the dashed curve.  Panel (e) shows the spurious fraction versus 
detection S/N.  In both (d) and (e), the black and yellow curves are derived by requiring the matched input and output sources to have flux density 
ratios less than 2 and 4, respectively.  The difference between the yellow and black curves shows that the required correction for spurious sources 
sensitively depends on the details of the matching between the input and output source lists.  Panel (f) shows the fraction of observed sources that are blends 
of multiple sources.  The three curves represent the cases where the flux densities of the blended sources in the input list are at least 1/2, 1/4, and 1/8 
of the flux densities of the detected sources. The 1/4 curve best represents the multiple fraction to be found in
sensitive interferometric followup, if the sources are not clustered at $\sim10\arcsec$ scales.
\label{figA1}} 
\end{figure*}

To estimate flux boosting caused by the Eddington bias and faint confusing sources, 
we matched sources in the input and output catalogs of our simulations. For each output source,
we searched for input sources within a radius of a beam HWHM.  We consider that we have a match when
the flux densities of the input and output sources are within a factor of two of each other.  
This ensures that the input and output sources have similar flux densities, since for any output
source, it would be possible to associate it with an arbitrarily faint nearby input source, given that faint sources have 
much higher spatial density (we will come back and discuss the choice of this factor of two at the end of this section).
When there are multiple input sources meeting the above distance and flux ratio criteria, the brightest one is considered as the match.
We started from the brightest output source and moved down the list.  The matched input sources 
were not considered in the subsequent searches.  We repeated this for the 200 realizations.  
The output-to-input flux ratio of each matched pair is the flux boosting factor.
Fig~\ref{figA1} shows the flux boosting factor as a function of the output (observed) source flux density (panel a) and the input (intrinsic) source flux density (panel b).
Overall, the required correction for measured flux densities is around 20\% in panel (a).  
Panel (b) show that sources as faint as about 2 mJy (intrinsic) can be strongly boosted into our 4-$\sigma$ detection 
range at $\gtrsim4$ mJy, but with very low detection completeness (panel c).
The flux boosting factor in (b) is artificially capped at 2.0 for the faintest sources, because we require the flux densities of the input and output sources
to be within this factor. However, the curves do not suggest a much higher boosting factor if we remove this requirement.

To estimate the completeness and spurious fraction, we only considered input sources that have a chance of being detected, given
the noise levels at their locations.  This is necessary because our map has a highly non-uniform sensitivity distribution.
To perform this estimate, we used the $+1 \sigma$ flux boosting factor in Fig.~\ref{figA1} (a) to boost the flux densities
in the input source list.  Only those sources with boosted flux densities greater than 3.5 times and 4 times the local r.m.s.\ noise are considered,
corresponding to 3.5~$\sigma$ and 4~$\sigma$ detection thresholds.
We then matched the sources in the input and output catalogs using the same criteria as for the flux boosting estimate.
An output source without a match is considered as spurious (although its flux may still be real, contributed by multiple
fainter sources).  An input source without a match is considered as undetected (although it may contribute flux to a nearby source).
The detection completeness calculated this way is presented in Fig.~\ref{figA1} (c), for sources detected at $>4~\sigma$ and
$>3.5~\sigma$.   The spurious rate is presented in Fig.~\ref{figA1} (d) and (e) (black curves), as functions of the output (measured) flux density and
detection S/N, respectively.  

In Fig.~\ref{figA1} (e), we see that the spurious fraction for $\sim3.5~\sigma$ sources is approximately 50\%.
Even at $\sim4~\sigma$, the fraction is approximately 30\%. Both values appear somewhat high.
However, sources detected with such low S/N only comprise a small fraction in our sample.  
Fig.~\ref{figA1} (d) better reflects the overall spurious fraction, as here sources detected at different S/N levels
are mixed according to the realistic counts and sensitivity distribution of the map.
In Fig.~\ref{figA1} (d), there is a small bump in spurious fractions between 10 and 20 mJy.  This is because some of these brighter
sources are detected in the outer area where noise increases rapidly. 
For $>4~\sigma$ detections, the spurious rate is below 15\% above 7 mJy and below about 20\% in nearly
the entire flux density range of interest.  If we lower the threshold to 3.5 $\sigma$, the completeness would increase by almost a factor of two
at the faint end of 3--4 mJy, but the spurious rate would also increase to 35\%.
Given this, it is not obvious to us whether we should favor 3.5-$\sigma$ detections over 4-$\sigma$ detections
for number count estimates.  Furthermore, because of the nature of our Eq.~\ref{eq_correction},
the choice of detection threshold will not change the corrected source counts; it will only slightly change the interpretation of
the intrinsic flux density range over which the counts are measured. 

We compare the above results with other SCUBA-2 450\,$\mu$m surveys and do not find major discrepancies. 
Our flux boosting factor of 20\% is comparable to  those in \citet{casey13} and \citet{chen13a}, but seems much higher than that in \citet[their figure~3]{zavala17}.
The very low flux boosting factor in Zavala et al.\ might be caused by the difference in how the simulations were conducted.
The behavior of the completeness curves in Fig.~\ref{figA1} (c) is qualitatively similar to those presented by other authors 
\citep{geach13,casey13,zavala17}.  The differences in sensitivities and mapping strategies prevent further quantitative comparison.
Our spurious rates of 20--40\% near the detection limit is also comparable to that in \citet{casey13}.

Finally, to obtain a rough idea of source blending, we estimated the fraction of output sources that consist of multiple blended faint sources.
We again conducted source matching with a search radius of half a beam FWHM.  However, this time, for each output source, we 
consider input sources that are brighter than 1/2, 1/4, and 1/8 times the output flux density.  Fig.~\ref{figA1} (f) shows the fraction of output sources that
contain contributions from more than one input source that meet the above flux ratio thresholds.  
The ``$>1/2$'' curve (solid) represents cases where sources of nearly equal brightness are blended and detected;
these are rare, and the blending fraction does not exceed 10\%.  The  ``$>1/4$'' curve (dashed) probably better mimics the
multiple fractions to be found in sensitive interferometric surveys, since the aimed detection S/N for the single-dish fluxes is typically between 10 and 20 and
a four times fainter companion can be detected at $>3~\sigma$.  This suggests that we will find roughly 10--20\% of our sources to be multiples 
if we conduct deep ALMA Band-9 imaging followup.  The majority of the multiple source are pairs, and blends of more than three sources only account for 6\% of them.
This predicted multiple fraction is much lower than the 35\%--60\% multiple fractions typically found in 
interferometric followup of 850\,$\mu$m SCUBA-2 and LABOCA sources \citep{barger12,hodge13,simpson15},
This is a consequence of the smaller SCUBA-2 beam at 450\,$\mu$m.
Nevertheless, the fraction of 10\% to 20\% is  comparable to the error in the amplitudes of our counts, meaning that its effect
is not negligible below 10 mJy.  The ``$>1/8$'' curve (dotted) steeply raises from above 20 mJy to below 10 mJy, but artificially 
saturates at 6~mJy because we do not include $<1$ mJy sources in our simulations.  

We point out that  the results presented in this section are highly sensitive to the details of the source matching criteria between the input and 
output lists.  For example, if we allow the flux density ratio between the matched input and output sources to be greater than two
(i.e., make easier to find a match), we can artificially decrease the spurious fraction, increase the flux boosting factor, and 
extend the derived intrinsic counts to fainter flux density limits.
To demonstrate this, we show the spurious fraction derived with input and output flux ratios relaxed to 4 in Fig.~\ref{figA1} (d) and (e) with yellow curves.
For 4-$\sigma$ detections, the spurious fraction is below 5\% over nearly the entire flux density range, but many of the added detections have 
flux densities that are dramatically boosted (more than a factor of two).  We stress again that we do not base our source count correction on the 
corrections of flux boosting, and completeness and spurious sources, because of all the above ambiguities and the 
arbitrariness in the choice of source matching criteria.

\section{Effect of Clustering}\label{sec_a2}

The results in the previous section are a strong indication that our counts will be biased if we do not properly take 
source blending into account when we correct the observed counts using Eq.~\ref{eq_correction_old}.  Our Eq.~\ref{eq_correction} does not suffer from this complication. 
On the other hand, our simulations (as well as similar simulations in the literature) assume a random (unclustered) spatial distribution.
Our counts will still be biased if the clustering of the 450\,$\mu$m sources at the $10\arcsec$ scale (20 comoving kpc) is strong enough 
to alter the blending fraction.  The simulations of \citet{cowley15} and \citet{bethermin17} show that clustering is not likely to have a strong effect 
on the observed counts for our beam size.  The ALMA observations of 870 $\mu$m single-dish selected sources of \citet{hodge13} seem to also support 
weak or no clustering at such small scales.  Future ALMA observations of 450\,$\mu$m sources are needed to further test the small-scale clustering
in the 450\,$\mu$m population. In the mean while, we use the simulations of \citet{bethermin17}, which includes clustering effects, to test whether clustering would bias our results.

\begin{figure*}[t!]
\epsscale{1.0}
\plotone{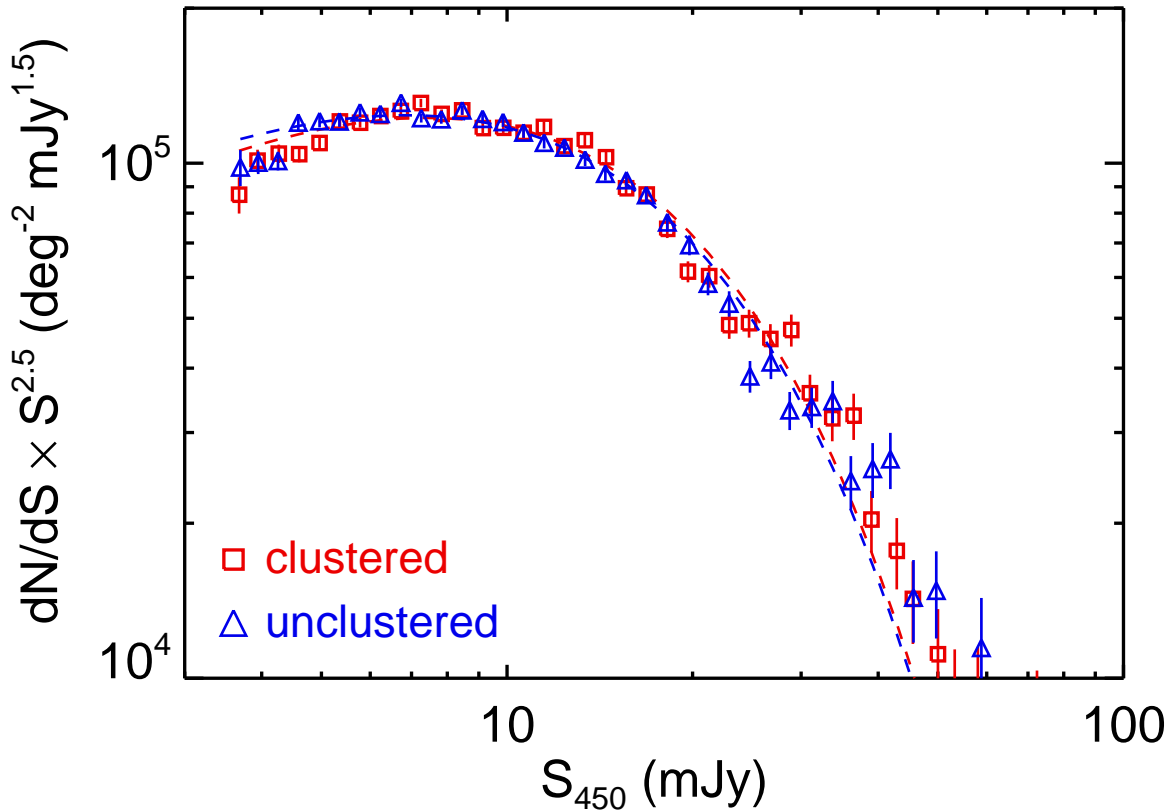}
\caption{Raw SCUBA-2 450\,$\mu$m counts derived with the simulations of \citet{bethermin17}.
The red symbols are the results with source clustering and the blue symbols without clustering.
The dashed curves are Schechter fits to the counts. 
\label{figA2}} 
\end{figure*}

We used the 2 deg$^2$ lightcone provided by \citet{bethermin17} and the associated 450\,$\mu$m fluxes of the simulated sources.
The general procedure is identical to our simulations described in Section~\ref{sec_counts}, except that we use sources in \citet{bethermin17} for the input.
We convolved the simulated sources with our SCUBA-2 PSF, and randomly pick a location in the 2 deg$^2$ region to simulate an observed
image.  This is repeated 100 times each for the positive and negative true noise map.  
We performed source extraction over the 200 simulated maps and derived the raw (observed) source counts.
The results represent the observed counts where sources are clustered, and are shown with red symbols and curve in Fig.~\ref{figA2}.
We then randomized the positions of the simulated sources in the 2 deg$^2$ region, and repeated the procedure.
The results represent the observed counts where sources are not clustered, but with identical intrinsic counts, and are shown with
blue symbols and curve in Fig.~\ref{figA2}. 
With the fitted Schechter functions (dashed curves), we find that at 5, 10, and 20 mJy, the counts with clustering are
higher than the unclustered cases by $-3.2\%$, +0.6\%, and +3.6\%, respectively.
These are essentially negligible, and do not suggest a systematic bias caused by source clustering.
By observing the data points in Fig.~\ref{figA2}, it is apparent that the accuracy of this test is limited by the size of the lightcone as well
as the area coverage of our observations.  We thus confirm that our counts are not biased by clustering, thanks to the small beam size of SCUBA-2 at 450\,$\mu$m.

\begin{figure*}[t!]
\epsscale{1.0}
\plotone{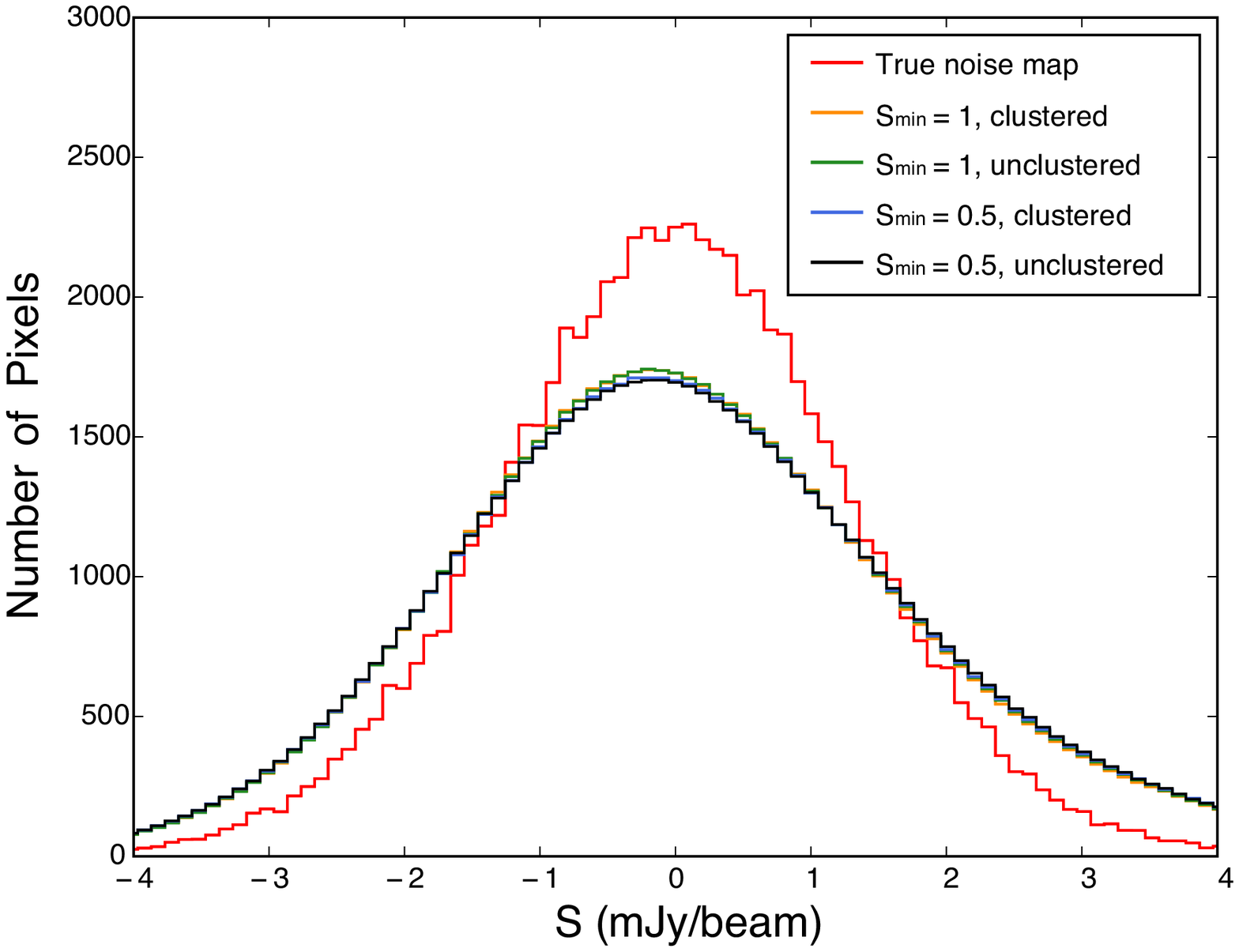}
\caption{Distributions of pixel brightness in the residual maps from the simulations using the lightcone of \citet{bethermin17} and the true noise map (red).
For both $S_{\rm min}=0.5$ and 1.0, the distributions for the clustered and unclustered cases are nearly identical. The differences in the log-likelihood
are also negligible.
\label{figA3}} 
\end{figure*}

Next, we test whether clustering would bias the fluctuation analyses for the faint end.  We again use the simulations of \citet{bethermin17}, with the 
original simulated source positions for the clustered case and randomized source positions for the unclustered case.
The procedure is identical to that described in Section~\ref{sec_mle}.  For simplicity, we tested the cases with faint-end cutoffs of $S_{\rm min}=0.5$ and 1.0 mJy.
We created 400 simulated maps for each case, and calculate the log-likelihood against the observed residual map.
The pixel brightness distributions of the simulated residual maps are shown in Fig.~\ref{figA3}.  For $S_{\rm min}=1.0$ mJy,
the dispersions of the log-likelihoods of the 400 simulated maps are 104 (clustered) and 73 (unclustered), while the mean difference between the clustered and unclustered 
log-likelihoods is 37.  For $S_{\rm min}=0.5$ mJy, the dispersions of the log-likelihoods are 109 (clustered) and 89 (unclustered), while the mean 
difference between the clustered and unclustered log-likelihoods is 25.  In both cases, the difference induced by source clustering is much smaller than
the variance in the ensemble of simulated images.  This can also be seen in Fig.~\ref{figA3}, where the clustered and unclustered curves
essentially overlap, and the difference caused by clustering is much smaller than changing the cutoff flux.  Based on these, we conclude that
our fluctuation analyses are not likely to be biased by source clustering at the scale of the SCUBA-2 450\,$\mu$m beam.


\begin{thebibliography}{} 
\bibitem[Alexander et al.(2005)]{alexander05}
	Alexander, D.\ M., Bauer, F.\ E., Chapman, S.\ C., Smail, I., Blain, A.\ W., Brandt, W.\ N., Ivison R.\ J., 2005, \apj, 632, 736
\bibitem[Almaini et al.(2003)]{almaini03}
	Almaini, O., Scott, S.\ E., Dunlop, J.\ S., et al.\ 2003, \mnras, 338, 303
\bibitem[Aravena et al.(2016)]{aravena16}
	Aravena, M., Decarli, R., Walter, F., et al.\ 2016, \apj, 833, 68
\bibitem[Asboth et al.(2016)]{asboth16}
	Asboth, V., Conley, A., Sayers, J., et al.\ 2016, \mnras, 462, 1989
\bibitem[Barger et al.(1998)]{barger98}
	Barger, A.\ J., Cowie, L.\ L., Sanders, D.\ B., Fulton, E., Taniguchi, Y., Sato, Y., Kawara, K., \& Okuda, H., 1998, \nat, 394, 248	
\bibitem[Barger, Cowie, \& Sanders(1999)]{barger99}
	Barger, A.\ J., Cowie, L.\ L., \& Sanders, D.\ B.\ 1999, \apjl, 518, L1 
\bibitem[Barger, Cowie, \& Richards(2000)]{barger00}
	Barger, A.\ J., Cowie, L.\ L., \& Richards, E.\ A.\ 2000, \aj, 119, 2092
\bibitem[Barger et al.(2012)]{barger12}
	Barger, A.\ J., Wang, W.-H., Cowie, L.\ L., et al.\ 2012, \apj, 761, 89
\bibitem[Barger et al.(2014)]{barger14}
	Barger, A.\ J., Cowie, L.\ L., Chen, C.-C., et al.\ 2014, \apj, 784, 9
\bibitem[Baugh et al.(2005)]{baugh05}
	Baugh, C.\ M., Lacey, C.\ G., Frenk, C.\ S., Granato, G.\ L., Silva, L., Bressan, A., Benson, A.\ J., \& Cole, S.\ 2005, \mnras, 356, 1191
\bibitem[B\'{e}thermin et al.(2012a)]{bethermin12a}
	B\'{e}thermin, M., Le Floc'h, E., Ilbert, O., et al.\ 2012a, \aap, 542, A58
\bibitem[B\'{e}thermin et al.(2012b)]{bethermin12b}
	B\'{e}thermin, M., Daddi, E., Magdis, G., et al.\  2012b, \apjl, 757, L23
\bibitem[B\'{e}thermin et al.(2017)]{bethermin17}
	B\'{e}thermin, M., Wu, H.-Y., Lagache, G., et al.\  2017, \aap, in press (arXiv:1703.08795)
\bibitem[Blain et al.(2002)]{blain02}
	Blain, A.\ W., Smail, I., Ivison, R.\ J., Kneib, J.-P., \& Frayer, D.\ T.\ 2002, PhR, 369, 111
\bibitem[Borys et al.(2003)]{borys03}
	Borys, C., Chapman, S., Helpern, M., \& Scott, D.\ 2003, \mnras, 344, 385
\bibitem[Bourne et al.(2017)]{bourne17}
	Bourne, N., Dunlop, J.\ S., Merlin, E., et al.\ 2017, \mnras, 467, 1360
\bibitem[Bouwens et al.(2015)]{bouwens15} 
	Bouwens, R.\ J., Illingworth, G.\ D., Oesch, P.\ A., et al.\ 2015, \apj, 803, 34 
\bibitem[Bouwens et al.(2017)]{bouwens17} 
	Bouwens, R.\ J., Oesch, P.\ A., Illingworth, G.\ D., Ellis, R.\ S., \& Stefanon, M.\ 2017, \apj, 843, 129 
\bibitem[Carniani et al.(2015)]{carniani15}
	Carniani, S., Maiolino, R., De Zotti, G., et al.\ 2015, \aap, 584, 78
\bibitem[Casey et al.(2013)]{casey13}
	Casey, C.\ M., Chen, C.-C., Cowie, L. L., et al.\ 2013, \mnras, 436, 1919
\bibitem[Casey, Narayanan, \& Cooray(2014)]{casey14}
	Casey, C.\ M., Narayanan, D., \& Cooray, A. 2014, PhR,  541, 45
\bibitem[Casey et al.(2015)]{casey15}
	Casey, C.\ M., Cooray, A., Capak, P., et al.\ 2015, \apjl, 808, L33
\bibitem[Chapin et al.(2009)]{chapin09}
	Chapin, E.\ L., Pope, A., Douglas, S., et al.\ 2009, \mnras, 398, 1793
\bibitem[Chapin et al.(2013)]{chapin13}
	Chapin, E.\ L., Berry, D.\ S., Gibb, A.\ G., et al.\ 2013, \mnras, 430, 2545
\bibitem[Chapman et al.(2003a)]{chapman03a}
   	Chapman, S.\ C., Blain, A.\ W., Ivison, R.\ J., \& Smail, I.\ R.\ 2003a, \nat, 422, 695
\bibitem[Chapman et al.(2003b)]{chapman03b}
   	Chapman, S.\ C., Windhorst, R., Odewahn, S., Yan, H., \& Conselice, C.\ 2003b, \apj, 599, 92
\bibitem[Chapman et al.(2005)]{chapman05}
	Chapman, S.\ C., Blain, A.\ W., Smail, I., \& Ivison, R.\ J.\ 2005, \apj, 622, 772
%\bibitem[Chary \& Elbaz(2001)]{chary01}
%	Chary, R., \& Elbaz, D. 2001, \apj, 556, 562
\bibitem[Chen et al.(2013a)]{chen13a}
	Chen, C.-C., Cowie, L.\ L., Barger, A.\ J., Casey, C.\ M., Lee, N., Sanders, D.\ B., 
	Wang, W.-H., \& Williams, J.\ P.\ 2013a, \apj, 762, 81
\bibitem[Chen et al.(2013b)]{chen13b}
	Chen, C.-C., Cowie, L.\ L., Barger, A.\ J., Casey, C.\ M., Lee, N., Sanders, D.\ B., 
	Wang, W.-H., \& Williams, J.\ P.\ 2013b, \apj, 776, 131
\bibitem[Chen et al.(2014)]{chen14}
	Chen, C.-C., Cowie, L.\ L., Barger, A.\ J., Wang, W.-H., \& Williams, J.\ P.\ 2014, \apj, 789, 12
\bibitem[Chen et al.(2015)]{chen15}
	Chen, C.-C., Smail, I., Swinbank, A.\ M., et al.\ 2015, \apj, 799, 194
\bibitem[Chen et al.(2016)]{chen16}
	Chen, C.-C., Smail, I., Ivison, R.\ J., et al.\ 2016, \apj, 820, 82
\bibitem[Chu et al.(2017)]{chu17}
	Chu, J.\ K., Sanders, D.\ B., Larson, K.\ L., et al.\ 2017, \apjs, 229, 25
\bibitem[Clements et al.(2010)]{clements10}
	Clements, D.\ L., Rigby, E., Maddox, S., et al.\ 2010, \aap, 518, L8
\bibitem[Condon(1974)]{condon74}
	Condon, J.\ J.\ 1974, \apj, 188, 279
\bibitem[Conselice, Chapman, \& Windhorst(2003)]{conselice03}
	Conselice, C.\ J., Chapman, S.\ C., \& Windhorst, R.\ A.\ 2003, \apjl, 596, L5
\bibitem[Coppin et al.(2006)]{coppin06}
	Coppin, K., Chapin, E.\ L., Mortier, A.\ M.\ J., et al.\ 2006, \mnras, 372, 1621
\bibitem[Cowie, Barger, \& Kneib(2002)]{cowie02}
	Cowie, L.\ L., Barger, A.\ J., \& Kneib, J.-P.\ 2002, \aj, 123, 2197
\bibitem[Cowie et al.(2017)]{cowie17}
	Cowie, L.\ L., Barger, A.\ J., Hsu, L.-Y., Chen, C.-C., Owen, F.\ N., \& Wang, W.-H.\ 2017, \apj, 837, 139
\bibitem[Cowley et al.(2015)]{cowley15}
	Cowley, W.\ I., Lacey, C.\ G., Baugh, C.\ M., \& Cole, S.\ 2015, \mnras, 446, 1784
\bibitem[Danielson et al.(2017)]{danielson17}
	Danielson, A.\ L.\ R., Swinbank, A.\ M., Smail, I., et al.\ 2017, \apj, 840, 78
\bibitem[Dannerbauer, Walter, \& Morrison(2008)]{dannerbauer08}
	Dannerbauer, H., Walter, F., \& Morrison, G.\ E.\ 2008, \apjl, 673, L127
\bibitem[Dempsey et al.(2012)]{dempsey12}
	Dempsey, J.\ T., Holland, W.\ S., Chrysostomou, A., et al.\ 2012, Proc.\ SPIE, 8452, 2
\bibitem[Dempsey et al.(2013)]{dempsey13}
	Dempsey, J.\ T., Friberg, P., Jenness, T., et al.\ 2013, \mnras, 430, 2534
\bibitem[Dunlop et al.(2017)]{dunlop17}
	Dunlop, J.\ S., McLure, R.\ J., Biggs, A.\ D., et al.\ 2017, \mnras, 466, 861
\bibitem[Eales et al.(1999)]{eales99}
	Eales, S., Lilly, S., Gear, W., et al.\ 1999, \apj, 515, 518
\bibitem[Fixsen et al.(1998)]{fixsen98}
	Fixsen, D.\ J., Dwek, E., Mather, J.\ C., Bennett, C.\ L., \& Shafer, R.\ A.\ 1998, \apj, 508, 123
\bibitem[Fujimoto et al.(2016)]{fujimoto16}
	Fujimoto, S., Ouchi, M., Ono, Y., Shibuya, T., Ishigaki, M., Nagai, H, \& Momose, R.\ 2016, \apjs, 222, 1 (F16)
\bibitem[Furusawa et al.(2008)]{furusawa08}
	Furusawa, H., Kosugi, G., Akiyama, M., et al.\ 2008, \apjs, 176, 1 
\bibitem[Geach et al.(2013)]{geach13}
	Geach, J.\ E., Chapin, E.\ L., Coppin, K.\ E.\ K., et al.\ 2013, \mnras, 432, 53
\bibitem[Geach et al.(2017)]{geach17}
	Geach, J.\ E., Dunlop, J.\ S., Halpern, M., et al.\ 2017, \mnras, 465, 1789
\bibitem[Gispert, Lagache, \& Puget (2000)]{gispert00}
	Gispert, R., Lagache, G., \& Puget, J.\ L.\ 2000, \aap, 360, 1
\bibitem[Glenn et al.(2010)]{glenn10}
	Glenn, J., Conley, A., B\'{e}thermin, M., et al.\ 2010, \mnras, 409, 109
\bibitem[Greve et al.(2004)]{greve04}
	Greve, T.\ R., Ivison, R.\ J., Bertoldi, F., et al.\ 2004, \mnras, 354, 779
\bibitem[Grogin et al.(2011)]{grogin11}
	Grogin, N., Kocevski, D., Faber, S.\ M., et al.\ 2011, \apjs, 197, 35
\bibitem[Hatsukade et al.(2011)]{hatsukade11}
	Hatsukade, B., Kohno, K., Aretxaga, I., et al.\ 2011, \mnras, 411, 102
\bibitem[Hatsukade et al.(2013)]{hatsukade13}
	Hatsukade, B., Ohta, K., Seko, A., Yabe, K., \& Akiyama, M.\ 2013, \apjl, 769, L27
\bibitem[Hatsukade et al.(2016)]{hatsukade16}
	Hatsukade, B., Kohno, K., Umehata, H., et al.\ 2016, \pasj, 68, 36 
\bibitem[Hayward et al.(2011)]{hayward11}
	Hayward, C.\ C.,  Kere\v{s}, D., Jonsson, P., et al.\ 2011, \apj, 743, 159
\bibitem[Hayward et al.(2012)]{hayward12}
	Hayward, C.\ C., Jonsson, P., Kere\v{s}, D., et al.\ 2012, \mnras, 424, 951
\bibitem[Hayward et al.(2013)]{hayward13}
	Hayward, C.\ C., Narayanan, D.., Kere\v{s}, D., Jonsson, P., Hopkins, P.\ F., Cox, T.\ J., \& Hernquist, L.\ 2013, \mnras, 428, 2529
\bibitem[Hodge et al.(2013)]{hodge13}
	Hodge, J.\ A., Karim, A., Smail, I., et al.\ 2013, \apj, 768, 91
\bibitem[Hodge et al.(2016)]{hodge16}
	Hodge, J.\ A., Swinbank, A.\ M., Simpson, J.\ M., et al.\ 2016, \apj, 833, 103
\bibitem[H\"{o}gbom(1974)]{hogbom74}
	H\"{o}gbom, J.\ A.\ 1974, \aaps, 15, 417
\bibitem[Holland et al.(1999)]{holland99}
	Holland, W.\ S., Robson, E.\ I., Gear, W.\ K., et al.\ 1999, \mnras, 303, 659
\bibitem[Holland et al.(2013)]{holland13}
	Holland, W.\ S., Bintley, D., Chapin, E.\ L., et al.\ 1999, \mnras, 430, 2513
\bibitem[Hsu et al.(2016)]{hsu16}
	Hsu, L.-Y., Cowie, L.\ L., Chen, C.-C., Barger, A.\ J., \& Wang, W.-H.\ 2016, \apj, 829, 25
\bibitem[Hughes et al.(1998)]{hughes98}
	Hughes, D.\ H., Serjeant, S., Dunlop, J., et al.\ 1998, \nat, 394, 241
\bibitem[Hung et al.(2016)]{hung16}
	Hung, C.-L., Caser, C.\ M., Chinag, Y.-K., et al.\ 2016, \apj, 826, 130
\bibitem[Ivison et al.(2016)]{ivison16}
	Ivison, R.\ J., Lewis, A.\ J.\ R., Wei\ss, A., et al.\ 2016, \apj, 832, 78
\bibitem[Jenness et al.(2008)]{jenness08}
	Jenness, T., Cavanagh, B., Economou, F., \& Berry, D.\ S.\ 2008, in ASP Conf.\ Ser.\ 394,
	Astronomical Data Analysis Software and Systems XVII, 
	ed.\ R.\ W.\ Argyle, P.\ S.\ Bunclark, \& J.\ R.\ Lewis. (San Francisco, CA: ASP), 	565
\bibitem[Jenness et al.(2011)]{jenness11}
	Jenness, T., Berry, D., Chapin, E., et al. 2011, in ASP Conf.\ Ser.\ 442, 
	Astronomical Data Analysis Software and Systems XX, ed.\ I.\ N.\ Evans et al. (San Francisco, CA: ASP), 281
\bibitem[Johansson, Sigurdarson, \& Horellou(2011)]{johansson11}
	Johansson, D., Sigurdarson, H., \& Horellou C.\ 2011, \aap, 527, 117
\bibitem[Karim et al.(2013)]{karim13}
	Karim, A., Swinbank, A.\ M., Hodge, J.\ A., et al.\ 2013, \mnras, 432, 2
\bibitem[Knudsen, van der Werf, \& Kneib(2008)]{knudsen08} 
	Knudsen, K.\ K., van der Werf, P.\ P., Kneib, J.-P.\ 2008, \mnras, 384, 1611
\bibitem[Koekemoer et al.(2011)]{koekemoer11}
	Koekemoer, A.\ M., Faber, S., Ferguson, H.\ et al. 2011, \apjs, 197, 36
\bibitem[Lacey et al.(2016)]{lacey16}
	Lacey, C.\ G., Baugh, C.\ M., Carlos S., F., et al.\ 2016, \mnras, 462, 3854
\bibitem[Laird et al.(2010)]{laird10}
	Laird, E.\ S., Nandra, K., Pope, A., \& Douglas, S.\ 2010, \mnras, 401, 2763
%\bibitem[Lenz, Hensley, \& Dor\'{e}(2017)]{lenz17}
%	Lenz, D., Hensley, B.\ S., \& Dor\'{e}, O.\ 2017, arXiv:1706.00011
\bibitem[Maloney et al.(2005)]{maloney05}
	Maloney, P.\ R., Glenn, J., Aguirre, J.\ E., et al.\ 2005, \apj, 635, 1044
\bibitem[Men\'{e}ndez-Delmestre et al.(2009)]{menendez09}
	Men\'{e}ndez-Delmestre, K., Blain, A.\ W., Smail, I., et al.\ 2009, \apj, 699, 667
\bibitem[Micha{\l}owski, Hjorth, \& Watson(2010)]{michalowski10}
	Micha{\l}owski, M., Hjorth, J., \& Watson, D.\ 2010, \aap, 514, 67
\bibitem[Micha{\l}owski et al.(2017)]{michalowski17}
	Micha{\l}owski, M., Dunlop, J.\ S., Koprowski, M.\ P.\ 2017, \mnras, 469, 492
\bibitem[Narayanan et al.(2010)]{narayanan10}
	Narayanan, D., Hayward, C.\ C., Cox, T.\ J.\ et al.\ 2010, \mnras, 401, 1613
\bibitem[Narayanan et al.(2015)]{narayanan15}
	Narayanan, D., Turk, M., Feldmann R.\ et al.\ 2015, \nat, 525, 496
\bibitem[Oliver et al.(2010)]{oliver10}
	Oliver, S.\ J., Wang, L., Smith, A.\ J., et al.\ 2010, \aap, 518, L21
\bibitem[Ono et al.(2014)]{ono14}
	Ono, Y., Ouchi, M., Kurono, Y., \& Momose, R.\ 2014, \apj, 795, 5
\bibitem[Oteo et al.(2016)]{oteo16}
	Oteo, I., Zwaan, M.\ A., Ivison, R.\ J., Smail, I., \& Biggs, A.\ D.\ 2016, \apj, 822, 36
\bibitem[Patanchon et al.(2009)]{patanchon09}
	Patanchon, G., Ade, P.\ R., Bock, J.\ J., et al.\ 2009, \apj, 707, 1750
%\bibitem[Planck Collaboration et al.(2014 XXX)]{planckxxx}
%	Planck Collaboration XXX. 2014, \aap, 571, A30
%\bibitem[Planck Collaboration et al.(2016 VIII)]{planckviii}
%	Planck Collaboration VIII. 2016, \aap, 594, A8
\bibitem[Pope et al.(2008)]{pope08}
	Pope, A., Chary, R.-R., Alexander, D.\ M., et al.\ 2008, \apj, 675, 1171
\bibitem[Puget et al.(1996)]{puget96}
	Puget, J.-L., Abergel, A., Bernard, J.-P., Boulanger, F., Burton, W.\ B., Desert, F.-X., \& Hartmann, D.\ 1996, \aap, 308, L5
\bibitem[Reddy \& Steidel(2009)]{reddy09} 
	Reddy, N.\ A., \& Steidel, C.\ C.\ 2009, \apj, 692, 778 
\bibitem[Roseboom et al.(2013)]{roseboom13}
	Roseboom, I.\ G., Dunlop, J.\ S., Cirasuolo, M., et al.\ 2013, \mnras, 436, 430
\bibitem[Sawicki \& Thompson(2006)]{sawicki06} 
	Sawicki, M., \& Thompson, D.\ 2006, \apj, 642, 653 
\bibitem[Serjeant et al.(2003)]{serjeant03}
	Serjeant, S., et al.\ 2003, \mnras, 344, 887
\bibitem[Scott et al.(2010)]{scott10}
	Scott, K.\ S., Yun, M.\ S., Wilson, G.\ W., et al.\ 2010, \mnras, 405, 2260
\bibitem[Scoville et al.(2007)]{scoville07}
	Scoville, N., Aussel, H., Brusa, M., et al.\ 2007, \apjs, 172, 1
\bibitem[Shu et al.(2016)]{shu16}
	Shu, X.\ W., Elbaz, E., Bourne, N., et al.\ 2016, \apjs, 222, 4
\bibitem[Simpson et al.(2014)]{simpson14}
	Simpson, J.\ M., Swinbank, A.\ M., Smail, I., et al.\ 2014, \apj, 788, 125
\bibitem[Simpson et al.(2015a)]{simpson15}
	Simpson, J.\ M., Smail, I., Swinbank, A.\ M., et al.\ 2015a, \apj, 799, 81
\bibitem[Simpson et al.(2015b)]{simpson15b}
	Simpson, J.\ M., Smail, I., Swinbank, A.\ M., et al.\ 2015b, \apj, 807, 128
\bibitem[Simpson et al.(2017)]{simpson17}
	Simpson, J.\ M., Smail, I., Wang, W.-H., et al.\ 2017, \apjl, 884, L10
\bibitem[Smail, Ivison, \& Blain(1997)]{smail97}
	Smail, I., Ivison, R.\ J., Blain, A.\ W., 1997, \apjl, 490, L5
\bibitem[Smail et al.(2002)]{smail02}
	Smail, I., Ivison, R.\ J., Blain, A.\ W., \& Kneib, J.-P.\ 2002, \mnras, 331, 495
%\bibitem[Smol\v{c}i\'{c} et al.(2012)]{smolcic12}
%	Smol\v{c}i\'{c}, V., Aravena, M., Navarrete, F., et al.\ 2012, \aap, 548, 4
\bibitem[Smol\v{c}i\'{c} et al.(2017)]{smolcic17}
	Smol\v{c}i\'{c}, V., Novak, M., Bondi, M., et al.\ 2017, \aap, in press (arXiv:1703.09713)
\bibitem[Steidel et al.(1999)]{steidel99} 
	Steidel, C.\ C., Adelberger, K.\ L., Giavalisco, M., Dickinson, M., \& Pettini, M.\ 1999, \apj, 519, 1 
\bibitem[Stetson(1987)]{stetson87}
	Stetson, P.\ B.\ 1987, \pasp, 99, 191
\bibitem[Strandet et al.(2016)]{strandet16}
	Strandet, M.\ L., Wei\ss, A., Vieira, J.\ D., et al.\ 2016, \apj, 822, 80
\bibitem[Swinbank et al.(2010)]{swinbank10}
	Swinbank, A.\ M., Smail, I., Chapman, S.\ C.\ et al.\ 2010, \mnras, 405, 234
\bibitem[Swinbank et al.(2011)]{swinbank11}
	Swinbank, A.\ M., Papadopoulos, P.\ P., Cox, P., et al.\ 2011, \apj, 742, 11
\bibitem[Swinbank et al.(2014)]{swinbank14}
	Swinbank, A.\ M., Simpson, J.\ M., Smail, I., et al.\ 2014, \mnras, 438, 1267
\bibitem[Umehata et al.(2015)]{umehata15}
	Umehata, H., Tamura, Y., Kohno, K., et al.\ 2015, \apjl, 815, L8
\bibitem[Valiante et al.(2007)]{valiante07}
	Valiante, E., Lutz, D., Sturm, E., Genzel, R., Tacconi, L.\ J., Lehnert, M.\ D., \& Baker, A.\ J.\ 2007, \apj, 660, 1060
\bibitem[Valiante et al.(2009)]{valiante09}
	Valiante, E., Lutz, D., Sturm, E., Genzel, R., \& Chapin, E.\ L.\ 2009, \apj, 701, 1814
\bibitem[Valiante et al.(2016)]{valiante16}
	Valiante, E., Smith, M.\ W.\ L., Eales, S., et al.\ 2016, \mnras, 462, 3146
\bibitem[Vernstrom et al.(2014)]{vernstrom14}	
	Vernstrom, T., Scott, D., Wall, J.\ V., et al.\ 2014, \mnras, 440, 2791
\bibitem[Wardlow et al.(2011)]{wardlow11}
	Wardlow, J.\ L., Smail, I., Coppin, K.\ E.\ K., et al.\ 2011, \mnras, 415, 1479
\bibitem[Wang, Cowie, \& Barger(2004)]{wang04}
	Wang, W.-H., Cowie, L.\ L., \& Barger, A.\ J.\ 2004, \apj, 613, 655
\bibitem[Wang et al.(2007)]{wang07}
	Wang, W.-H., Cowie, L.\ L., van Saders, J., Barger, A.\ J., \& Williams, J.\ P.\ 2007, \apjl, 670, L89
\bibitem[Wang, Cowie, \& Barger(2006)]{wang06}
	Wang, W.-H., Cowie, L.\ L., \& Barger, A.\ J.\ 2006, \apj, 647, 74
\bibitem[Wang et al.(2011)]{wang11}
	Wang, W.-H., Cowie, L.\ L., Barger, A.\ J., Williams, J.\ P.\ 2011, \apjl, 726, L18
\bibitem[Wang, Kohno, et al.(2016)]{wang16b}
	Wang, W.-H., Kohno, K., Hatsukade, B., et al.\ 2016, \apj, 833, 195
\bibitem[Wang et al.(2013)]{wang13}
	Wang, S.\ X., Brandt, W.\ N., Luo, B., et al.\ 2013, \apj, 778, 179	
\bibitem[Wang, Elbaz, et al.(2016)]{wang16}
	Wang, T., Elbaz, D., Daddi, E., et al.\ 2016, \apj, 828, 56
\bibitem[Wei\ss~et al.(2009)]{weiss09}
	Wei\ss, A., Kov\'{a}cs, A., Coppin, K., et al.\ 2009, \apj, 707, 1201
\bibitem[Wilkinson et al.(2017)]{wilkinson17}
	Wilkinson, A., Omar, A., Chen, C.-C., et al.\ 2017, \mnras, 464, 1380
\bibitem[Younger et al.(2007)]{younger07}
	Younger, J.\ D., Giovanni, F., Huang, J.-S., et al.\ 2007, \apj, 671, 1531
\bibitem[Zavala et al.(2014)]{zavala14}
	Zavala, J.\ A., Aretxaga, I., \& Hughes, D.\ H.\ 2014, \mnras, 443, 2384
\bibitem[Zavala et al.(2017)]{zavala17}
	Zavala, J.\ A., Aretxaga, I., Geach, J.\ E., et al.\ 2017, \mnras, 464, 3369





\end{thebibliography}
\end{document}